\documentclass[english]{scrartcl}
\usepackage[latin9]{inputenc}
\usepackage{geometry}
\usepackage{babel}
\usepackage{verbatim}
\usepackage{float}
\usepackage{enumitem}
\usepackage{bm}
\usepackage{amsmath}
\usepackage{amsthm}
\usepackage{amssymb}
\usepackage{graphicx}
\usepackage{esint}

\usepackage{xcolor}
\definecolor{blue}{HTML}{1F77B4}
\definecolor{orange}{HTML}{FF7F0E}
\definecolor{green}{HTML}{2CA02C}
\definecolor{red}{HTML}{D62728}
\definecolor{purple}{HTML}{9467BD}
\definecolor{brown}{HTML}{8C564B}
\definecolor{pink}{HTML}{E377C2}
\definecolor{grey}{HTML}{7F7F7F}
\definecolor{yellow}{HTML}{BCBD22}
\definecolor{cyan}{HTML}{17BECF}
\definecolor{turquoise}{HTML}{3FE0D0}

\usepackage{natbib}
\usepackage[unicode=true,pdfusetitle,
 bookmarks=true,bookmarksnumbered=true,bookmarksopen=true,bookmarksopenlevel=1,
 breaklinks=true,pdfborder={0 0 1},backref=false,colorlinks=true]
 {hyperref}
\hypersetup{
 linkcolor=blue, citecolor=blue, urlcolor=blue, filecolor=blue,pdfpagelayout=OneColumn, pdfnewwindow=true,pdfstartview=XYZ, plainpages=false, pdfpagelabels, hyperindex=true}

\usepackage {graphicx}
\usepackage{epsfig,amsmath,latexsym,amssymb}
\usepackage{bm}
\usepackage{booktabs}
\usepackage{multirow}
\usepackage{caption}
\usepackage{algorithm}
\usepackage{algpseudocode}
\usepackage{comment}

\newtheorem{remark}{Remark}[section]
\newtheorem{lem}{\protect\lemmaname}[section]
\providecommand{\lemmaname}{Lemma}

\newcommand{\E}{\mathbb{E}}
\newcommand{\Var}{\mathbb{V}\hspace{-0.08334em}\mathrm{ar}}

\usepackage{orcidlink}
\definecolor{orcidlogocol}{HTML}{2E7E9F} 
\usepackage{authblk}

\usepackage{enumitem}
\setlist{leftmargin=*, topsep=0.5em, parsep=0pt, itemsep=1em, labelindent=0pt, align=left}

\addtokomafont{title}{}

\begin{document}

\title{Deep Least Squares Monte Carlo methods for the valuation of variable annuities with guarantees}

\author[1]{\orcidlinki{\textcolor{black}{Nicolas Langren\'e}}{0000-0001-7601-4618}}
\author[2]{Xiaolin Luo}
\author[2]{\orcidlinki{\textcolor{black}{Pavel V.~Shevchenko}}{0000-0001-8104-8716}\thanks{Corresponding author, pavel.shevchenko@mq.edu.au}}
\author[3]{\orcidlinki{\textcolor{black}{Ruiyi Zhang}}{0009-0005-9414-2871}}

 \affil[1]{\normalsize Guangdong Provincial/Zhuhai Key Laboratory of Interdisciplinary Research and Application for Data Science, Beijing Normal-Hong Kong Baptist University, Zhuhai, 519087, Guangdong, China}
 \affil[2]{\normalsize Actuarial Studies and Business Analytics, Macquarie University, Sydney, 2109, New South Wales, Australia}
 \affil[3]{\normalsize Department of Mathematics, Hong Kong University of Science and Technology, Hong Kong, 999077, Hong Kong}

\maketitle

\begin{abstract}
\noindent
In general, the pricing of variable annuities with guarantees can be done by solving the corresponding optimal stochastic control problem if the contract withdrawal strategy is assumed to be optimal.
This is typically solved as a dynamic programming problem using deterministic grid methods,
which become computationally infeasible for more than a few state variables. In such situations, one needs to rely on simulation methods. The least-squares
Monte Carlo (LSMC) method has become a popular simulation method for solving
optimal stochastic control problems in quantitative finance over the last decades. In principle, the LSMC, originally developed for pricing Bermudan options, cannot be used directly for pricing variable annuities without simplifying assumptions because the underlying state variables are affected by the control decisions.
This paper presents modifications of the LSMC algorithm that makes the pricing of general variable annuities feasible. For numerical illustrations, the pricing of variable annuities with guaranteed minimum withdrawal benefit under optimal withdrawal strategies is obtained with and without stochastic interest rates, using either polynomial regression or neural network regression in the LSMC algorithm. We found that the classical polynomial LSMC can give very accurate prices, at the cost of manual feature engineering, and with a standard deviation of the estimator that increases greatly when interest rates are made stochastic. By contrast, neural network LSMC gives slightly less accurate prices, requires more training time, but does not require manual feature engineering, and making interest rates stochastic makes no visible difference to its accuracy, suggesting a more stable and robust pricing performance of deep LSMC for higher-dimensional pricing problems. 

\end{abstract}

\section{Introduction}
\label{sec:introduction} 

The world population is ageing fast with life expectancy raising near 90 years in some countries.
Longevity risk, i.e., the risk of outliving one's savings, is becoming a critical concern for retirees.
Variable annuities (VA) with living and death benefit guarantees are financial
products that can help to manage this risk. These products provide savings protection while taking advantage of market growth at the same time.
VAs are typically classified based on the type of guarantees they provide. These include guaranteed minimum death benefit (GMDB), guaranteed minimum withdrawal benefit (GMWB), guaranteed minimum maturity benefit (GMMB), guaranteed minimum accumulation benefit (GMAB), and guaranteed minimum income benefit (GMIB) \citep{feng2022variable}. A good overview of VA products and the development of their market can be found in \citet{bauer2008universal}, \citet{ledlie2008annuities} and \citet{kalberer2009annuities}.
Insurers started to sell these types of products in the 1990s in United States. Later, these products became popular in Europe, UK and Japan as well. The market for VAs has become very large. In the United States, according to the LIMRA 
U.S. Individual Annuities Survey, annual sales of VAs between 2003 and 2014   ranged from \$129 to \$184 billion (peaking in 2007). Sales then declined to around \$100 billion per year by 2016 and remained near this level until 2023, before increasing to \$126.5 in 2024 and \$142.6 billion in 2025. It is also interesting to observe that sales of fixed annuities were below VA sales during 2003-2015, reaching as little as about half of VA sales in some years. However, this trend later reversed, with fixed annuity sales reaching \$307.6 in 2024 and \$321.5 billions in 2025, more than double the corresponding VA sales).

VAs with guarantees provide flexibility to the policyholder to withdraw funds from the contract, though some penalty may apply for deviating from the contractual rate. If the policyholder behaves passively and makes withdrawals at the contractual rate defined at the beginning of the contract, then the behaviour of the policyholder is called \emph{static}. In this case the paths of the wealth account can be simulated forward, and a standard Monte Carlo (MC) simulation method can be used for GMxB pricing. On the other hand, if the policyholder optimally decides the amount to withdraw at each withdrawal date, then the behaviour of the policyholder is called \emph{dynamic}. Under the optimal withdrawal strategy, the fair pricing of variable annuities with GMWB becomes a stochastic optimal control problem.

In the literature, popular numerical methods to price variable annuities include 
\begin{itemize}
\item Tree/lattice algorithms: GLWB in \citet{goudenege2016pricing}, GMWB in \citet{dong2019willow,goudenege2019pricing,fontana2023valuation}, GMWB and GMDB in \citet{goudenege2021gaussian}, GMMB and GMAB in \citet{martire2023surrender}, \item Numerical PDE methods: GMMB in \citet{kang2018optimal}, GLWB in \citet{goudenege2016pricing}, GMWB in \citet{gudkov2019pricing,goudenege2019pricing}, GMWB and GMDB in \citet{goudenege2021gaussian}, and

\item Numerical integration: GMWB in \citet{luo2015fast}, GMWB with surrender in \citet{luo2015variable}, GLWB in \citet{bacinello2022optimal}, GMAB in \citet{shevchenko2016unified,jeon2021pricing}, GMWDB in \citet{luo2015gmwd}, GMMB and GMAB in \citet{martire2023surrender}, including the Fourier cosine expansion method (GMWB in \citet{alonso2018pricing}, GMDB in \citet{yu2019valuing}, GMMB in \citet{kang2022valuation,zhong2023efficient,ai2023valuing,ai2024valuation}) and the frame project method (GMDB in \citet{kirkby2021equity}, GMAB and GMDB in \citet{kirkby2023valuation}, GMMB and GMDB in \citet{zhong2023valuation}).
\end{itemize}

All these numerical methods are expected to perform well when the number of stochastic state variables of the problem is low (one or two), and to become quickly infeasible when this number grows. When the problem involves multiple state variables, an attractive numerical method to solve discrete-time stochastic control problems with finite horizon is the Least-Squares Monte Carlo (LSMC, \citealt{carriere1996valuation}, \citealt{longstaff2001american}), due to the better convergence guarantees of Monte Carlo methods in high dimension. The LSMC algorithm was introduced to variable annuity pricing in \citet{bacinello2011variable} in the case of a semi-static strategy with only one surrender guarantee. This was later extended in \citet{huang2016regression} in the specific case of a dynamic strategy with bang-bang control (the holder of a GLWB can maximize the contract writer's losses by only performing non-withdrawal, withdrawal at exactly the contract rate, or full surrender). In such situation, the pricing problem can be formulated as an optimal switching problem, for which the LSMC algorithm is known to be easily applicable \citep{barreraesteve2006numerical,aid2014probabilistic,andersson2025effectiveness}.
Similarly, \citet{goudenege2019pricing} considered dynamic withdrawals under the simplifying assumption that the optimal withdraws should be multiples of the guaranteed rate.

In this paper, we revisit the problem of pricing variable annuities with embedded guarantees by LSMC, and generalize it to the case of fully dynamic optimal strategies with fully controlled underlying stochastic state variables, and no ex ante simplifications of the optimal dynamic strategy. 

Using the GMWB example, the standard LSMC algorithm cannot be applied directly, as the paths of the underlying wealth process are altered by the optimal cash withdrawals that should be found from the backward-in-time solution, meaning that the underlying wealth process cannot be initially simulated forward in time as expected from the LSMC procedure. To overcome this problem, we introduce the use of the control randomization technique \citep{kharroubi2014randomization,fuhrman2025randomization} to VA pricing, to extend LSMC to handle stochastic optimal control problems with controlled Markov processes. 

As emphasized in \citet{ludkovski2023unified} or \citet{andersson2025effectiveness} for example, the LSMC algorithm can perfectly work with any other type of regression than ordinary least squares regression (in which case ``Regression Monte Carlo'' is probably a more appropriate name for the resulting algorithm). In fact, while \citet{longstaff2001american} used OLS regression, the seminal article \citet{carriere1996valuation} used nonparametric regression instead (spline regression and local kernel regression). Since then, many other types of regression have been investigated for LSMC applied to optimal stopping problems (see \citet{ludkovski2023statistical} and references therein).

In the context of discrete-time, finite horizon stochastic control problems, estimating continuation values by neural network regression was investigated in \citet{hure2021convergence}. This approach has been applied to option hedging \citep{fecamp2021deep,bachouch2022numerical}, storage optimization \citep{bachouch2022numerical,warin2023reservoir}, portfolio optimization \citep{franco2022discrete,roch2023optimal}, utility maximization \citep{andreasson2024optimal,arandjelovic2026solving}, and, recently, to the pricing of convertible bonds with path-dependent provisions \citep{zhu2026deep}. In this article, we introduce this methodology to the pricing of variable annuities with embedded guarantees.

Finally, we also establish how variations of the methods can be implemented to easily obtain confidence intervals for the optimal contract value.

The paper is organized as follows. Section~\ref{sec:va_stochastic_control} details how to model the pricing of a variable annuity with Guaranteed Minimum Withdrawal Benefit (GMWB) under optimal dynamic withdrawals as a stochastic optimal control problem with controlled underlying stochastic factors. Section~\ref{sec:va_lsmc} describes how to extend the Least Squares Monte Carlo (LSMC) algorithm to solve such problem, and obtain confidence intervals for the optimal contract value. Section~\ref{sec:numerical_results} provides numerical experiments illustrating the accuracy of the method, and Section~\ref{sec:conclusion} concludes the paper.

\section{Pricing VA riders as a stochastic control problem\label{sec:va_stochastic_control}}

The specification details of VA riders vary across different companies offering these products and the results for specific GMxB riders presented in the academic literature often refer to different specifications. However, in general, the present value of the overall payoff of the VA contract with a guarantee can be written as

\begin{equation}\label{total_payoff_eq}
H_0({X},{\pi})= \beta_{0,N} R_N({X}_N)+\sum_{n=1}^{N-1} \beta_{0,n} R_n({X}_n,\pi_n),
\end{equation}
where $R_N(X_N)$ is the cashflow amount received by the policy holder at the contract maturity,
 $R_n({X}_n,\pi_n)$
is the cashflow amount received by the policyholder at time $t_n$, $\pi_n$ is the withdrawal amount from the contract account, and
$\beta_{i,j}$ is the discounting factor from $t_j$ to $t_i$ \begin{equation}
\beta_{i,j}=\exp\left(-\int_{t_i}^{t_j}r(t)dt\right),\;t_j>t_i,
\end{equation}
where $r(t)$ is the risk-free interest rate (possibly stochastic). That is, we consider the time discretization $0=t_0<t_1<\cdots<t_N=T$ corresponding to the contract withdrawal dates, where $t_0=0$ is today and $T$ is the contract maturity.
Here, ${X}=\{X_n\}_{n=0}^N$, and ${X}_n$ is the state variable vector (containing variables such as wealth account, guarantee account, etc.) before the withdrawal $\pi_n$ at time $t_n$ evolving as
\begin{equation}
\label{eq:transitionfunction}
{X}_{n+1} = \mathcal{T}_n \left ({X}_n, \pi_n, {Z}_{n+1} \right ),
\end{equation}
where ${Z}_1,...,{Z}_N$ are independent random disturbances (i.e., the state variable ${X}=\{X_n\}_{n=0}^N$ is a Markov
process). Under the assumption of no-arbitrage market with respect to the financial risk, the price of the VA with GMxB can be expressed
as an expectation with respect to the risk-neutral probability measure for the underlying
risky asset. Let $(\Omega,\mathcal{F},\mathbb{Q})$ be a probability space with sample space $\Omega$, filtration $\mathcal{F}=\{\mathcal{F}_t : t\ge 0\}$ and risk-neutral probability measure $\mathbb{Q}$. Then, the contract fair price under the given withdrawal strategy ${\pi}=(\pi_1,\ldots,\pi_{N-1})$, can be calculated as
\begin{equation}\label{GMWDB_general_eq}
V_0\left({X}_0\right)=\E\left[H_0({X},{\pi})|\mathcal{F}_0\right].
\end{equation}
Here, $\E[\cdot|\mathcal{F}_t]$ denotes an expectation under the risk-neutral probability measure ${\mathbb{Q}}$ with respect to the state vector ${X}$, conditional on information available at time $t$.

The insurer is collecting the fee $\alpha$ for the guarantee affecting the evolution of $X_t$ and the contract value $V_0(X_0;\alpha)$ depends on this fee rate. Then, the fair value of the fee $\alpha$ corresponds to the solution of the equation $V_0(X_0;\alpha)=W_0$, where $W_0$ is the full amount invested into the contract at $t_0$.

The withdrawal strategy $\pi$ can depend on the time and state
variables and is assumed to be given when the price of the contract
is calculated in \eqref{GMWDB_general_eq}. Different types of withdrawal strategies
exist. They may be classified as either \emph{static}, \emph{optimal}, or
\emph{sub-optimal}.
\begin{itemize}
\item \textbf{Static strategy.} Under this strategy, the policyholder decisions are deterministically determined at the beginning of the contract and do not depend on the evolution of the wealth and benefit base accounts. For example, the policyholder may withdraw at the contractual rate only.

\item \textbf{Optimal strategy.} Under the optimal withdrawal strategy, the decision on the withdrawal amount $\pi_n$ depends on the information available at time $t_n$, i.e. depends on the state variable $X_n$. The optimal strategy is calculated as
\begin{equation}\label{optimalstrategy_eq}
{\pi}^\ast({X})=\underset{{\pi\in\mathcal{A}}}{\mathrm{\arg\sup}}\ \E\!\left[H_0({X},{\pi})|\mathcal{F}_0\right],
\end{equation}
where the supremum is taken over all admissible strategies $\pi$ in space $\mathcal{A}$. Any other strategy ${\pi}({X})$ different from ${\pi}^\ast({X})$ is called \emph{sub-optimal} and leads to a smaller price.
\end{itemize}

Evaluating the contract \eqref{GMWDB_general_eq} under the
optimal withdrawal strategy \eqref{optimalstrategy_eq} is a standard
optimal stochastic control problem for a \emph{controlled Markov
process}. Note that, the control variable $\pi_n$ affects the transition law of the underlying state variable ${X}_t$ from $t_n$ to $t_{n+1}$ and thus the process is controlled. For a good textbook treatment of stochastic control
problems in finance, see \citet{bauerle2011markov}. This type of
problems can be solved recursively to find the contract value
$V_{n}(x)$ at $t_n$ when $X_n=x$ for $n=N-1,\ldots,0$ via the backward induction \emph{Bellman equation}
\begin{equation}\label{Bellman_eq}
V_{n}(x)=\sup_{\pi_n\in
\mathcal{A}_n}\left(R_n(x,\pi_n)+ \E\left[\beta_{n,n+1}
V_{{n+1}}(X_{n+1})\bigg| X_{n}=x;\pi_n \right]\right),
\end{equation}
starting from the final condition $V_N(x)=R_N(x)$. Note that the transition probability to reach state $X_{n+1}=x^\prime$ at time
$t_{n+1}$ if the withdrawal (\emph{action}) $\pi_n$ is applied in
the state $x$ at time $t_n$ depends on $\pi_n$.
Obviously, the above backward induction can also be used to calculate the fair contract price in the case of a static strategy $\pi$; in this case the space of admissible strategies $\mathcal{A}_n$ contains only one pre-defined value and $\sup(\cdot)$ becomes redundant.

\begin{remark} 
~
\begin{itemize}
    \item If the risk-free interest rate $r(t)$ is stochastic, then it can be convenient to use a change of num\'{e}raire technique. In particular, changing the num\'{e}raire from the money market account $M(t)=\exp(\int_0^t r(q)dq)$ to the bond price $\mathcal{P}(t_n,t_{n+1})$ at time $t_n$ with maturity $t_{n+1}$ will simplify calculations of the expectation in \eqref{Bellman_eq}, see \citet{shevchenko2017gmwb}.

\item If volatility, interest rates, or the force of mortality are stochastic, then these variables should be added to the state vector ${X}$ for valuation of the fair price expectation \eqref{GMWDB_general_eq}.
\end{itemize}
\end{remark}

\section{Numerical valuation of GMWB via LSMC\label{sec:va_lsmc}}
In general, solving the dynamic programming problem \eqref{Bellman_eq} needs to be done numerically, and can be very computationally intensive if a quadrature-based method is used for evaluating the expectation in \eqref{Bellman_eq} as the number of states, stochastic and control variables increases. The idea behind utilitising the LSMC method is to approximate the conditional expectation in \eqref{Bellman_eq},
\begin{equation}
\label{eq:conditionalexp}
\Phi_n(X_{n},\pi_n) = \E \left[ \beta_{n,n+1} V_{n+1}(X_{n+1}) | X_{n} ; \pi_n \right],
\end{equation}
by a regression scheme with independent variables $X_n$ and randomised $\pi_n$, and response variable $\beta V_{n+1}(X_{n+1})$. The approximation of the function is denoted as $\widehat{\Phi}_n$. Hereafter, for notational convenience we drop indexes for discounting factors.

\begin{remark}
~
\begin{itemize}
\item The pricing of VA with guarantees involves control variables (withdrawals) affecting the state variables. This is the so-called endogenous case and the conditional expectations $\Phi_t(X_{t},\pi_t)$ depends on the control $\pi_t$. The LSMC algorithm \citep{carriere1996valuation,longstaff2001american} was originally developed for pricing Bermudan options; this is an exogenous case where the state variables are not affected by the controls. In such case, only the regression approximation of $\Phi_t(X_{t}) = \E \left[ \beta V_{t+1}(X_{t+1}) | X_{t} \right]$ is required.
 \item For LSMC, the transition probability density function does not need to be known or evaluated. One only needs to be able to simulate the underlying state process.
\end{itemize} 
\end{remark}

\subsection{LSMC algorithms}
The LSMC proceeds first by simulation of the random state, control and disturbance variables for $X_t^m$, $\pi_t^m$, $m=1,...,M$, $t=0,...,T$ (forward simulation) as in Algorithm~\ref{simulation_algorithm}, where $Rand$ corresponds to random sampling from some distribution that could be designed for the specific problem. Then, the problem is solved with backward in time induction that can be accomplished using various algorithms. The most popular algorithms are: \emph{Realised Value} Algorithm~\ref{montecarlo_algorithm_realized} and \emph{Regression Surface} Algorithm~\ref{montecarlo_algorithm_regression} implemented in \emph{Regress Now} fashion. Both can be implemented in \emph{Regress Later} fashion; see Algorithms \ref{montecarlo_algorithm_realized_later} and \ref{algorithm_RegressionSurfaceRegressLater}.
Typical implementations of LSMC are based on the ordinary least squares (OLS) regression, which can be replaced with nonlinear regression such as, e.g. neural network regression used in our study. 

For \emph{Regress Now} algorithms, at each point in time $t<T$, the value function is approximated using the least-squares regression
\begin{equation}
\begin{split}
&\beta V_{n+1}(X^{m}_{n+1})=f_{\theta_n}(X^{m}_n,\pi^m_n)+\epsilon^{m}_n, \\ &\epsilon^{m}_n\overset{iid}{\sim} F_n(\cdot),\;\;\E[\epsilon^{m}_n]=0,\;\;\Var[\epsilon^{m}_n]=\sigma_n^2,\;\;m = 1,...,M,
\end{split}
\end{equation}
where $f_{\theta_n}(X^{m}_n,\pi^m_n)$ is a parameterised regression function (parameterised by a vector $\theta_n$). In the case of the OLS regression, then $f_{\theta_n}(X^{m}_n,\pi^m_n)=\bm{\Lambda}_n' \mathbf{L}(X^{m}_n,\pi^m_n)$,
where $\mathbf{L}(X^m_n,\pi^m_n)$ is a vector of basis functions and $\mathbf{\Lambda}_n$ the regression coefficient vector. In our study we consider polynomial basis functions and a deep neural network parameterisation for $f_\theta(X^{m}_n,\pi^m_n)$. Then, after the parameters $\hat\theta_n$ have been estimated,
\begin{equation}\label{unbiasedestimate_eq}
\widehat{{\Phi}}_n(X_n,\pi_n)={f}_{\hat\theta_n}(X_n,\pi_n). 
\end{equation}
For \emph{Regress Later}, the value function is approximated via the least-squares regression
\begin{equation}
\beta V_{n+1}(X^{m}_{n+1})=f_{\theta_{n+1}}(X^{m}_{n+1})+\epsilon^{m}_{n+1}, 
\end{equation}
where $f_{\theta_{n+1}}(X^{m}_{n+1})=\bm{\Lambda}_{n+1}' \mathbf{L}(X^{m}_{n+1})$ in the case of OLS,
and then the expectation 
\begin{equation}
\Phi_n(X_{n},\pi_n) = \E \left[ \beta_{n,n+1} V_{n+1}(X_{n+1}) | X_{n} ; \pi_n \right]
\end{equation}
is calculated. It can be calculated numerically via quadrature methods if the number of stochastic disturbances is one or two. If a polynomial basis is used and the moments of $X_{n+1}|X_n,\pi_n$ are known in closed form, then this expectation can be computed in closed form (see \ref{app:regress_later} for an example).

Sometimes, to avoid difficulties in the approximation of the value function, a transformation that has a similar shape as the value function can be applied on the value function in the above regression. Also, heteroscedasticity can be present in residuals, and then conditional variance can be modelled using another regression. These improvements of the standard LSMC algorithm are described in \citet{andreasson2022bias}.

Two types of LSMC implementations are \emph{Regression Surface} and \emph{Realised Value}. In the case of \emph{Regression Surface}, the value function is calculated in the backward recursion as
$$
\hat{V}_n(X_n)=R_n(X_n,\pi^\ast(X_n))+\Phi_n(X_n,\pi^\ast_n(X_n)).
$$
The \emph{Realised Value} method calculates the value function as
$$
\hat{V}_n(X_n)=R_n(X_n,\pi^\ast(X_n))+\beta\hat{V}_{n+1}(X_{n+1}),
$$
and requires the recomputation of the
sample paths from $t + 1$ to $T$ after each backwards-in-time iteration, as the optimal control affects the future state variables, hence changes the simulated paths \citep{langrene2015regression}.
Due to recalculation along the simulated path for each iteration, the computational complexity of the realised value method grows quadratically in the number of time steps compared with the linear growth for the regression surface method. However, the realised value method tends to accumulate much less regression errors over time and,
from experience, this method is much more suitable for problems prone to regression errors when the number of time periods increases.

We note that in the forward simulation Algorithm~\ref{simulation_algorithm}, the state variables and controls at time $t$ are sampled from problem specific distribution and then the state variables at $t+1$ are simulated according to the transition function (\ref{eq:transitionfunction}). That is, one does not need to simulate full trajectories of state variables according to the transition function (\ref{eq:transitionfunction}) for $n=0,...,N$. Instead, only samples from $t$ to $t+1$ are needed for regression in the corresponding backward solution. In some applications such as \citet{andreasson2024optimal, arandjelovic2026solving}, it appears critical; although in our illustrative GMWB examples, we did simulate full trajectories for $n=0,...,N$.

Finally, we want to mention that the use of post-decision variable may improve efficiency of the method if
 the transition equation~\eqref{eq:transitionfunction} can be separated into two transitions: 
\begin{equation}\label{equation-post-decision}
X_{t+1} = \widetilde{\mathcal{T}}_t(F(X_t, \pi_t), Z_{t+1}),
\end{equation} 
where the deterministic transition to the \emph{post-decision} variable $\widehat{X}_{t} = F(X_{t}, \pi_{t})$ precedes the transition $X_{t+1} = \widetilde{\mathcal{T}}_t(\hat{X}_{t}, Z_{t+1})$. 
This allows the conditional expectation \eqref{eq:conditionalexp} to be simplified as: 
\begin{equation} 
\mathbb{E}\big[\beta_{n,n+1}  V_{t+1}(X_{t+1})\, \big|\, X_{t}; \pi_{t} \big] = \mathbb{E}\big[\beta_{n,n+1}  V_{t+1}(X_{t+1})\, \big|\, \widehat{X}_{t} \big]. 
\end{equation} 
This method offers two main advantages: (1) dimension reduction in the covariates needed for the least-squares approximation of the conditional expectation, and (2) an increase in sampling efficiency by sampling only the post-decision states $\hat{X}_{t}$ rather than both $X_{t}$ and $\pi_{t}$. 


\begin{algorithm}[H]
\footnotesize
\caption{Forward simulation}
\label{simulation_algorithm}
\begin{algorithmic}[1]
\For{${n} = 0 \,\, \mathbf{to} \,\, N-1$}
\For{$m = 1 \,\, \mathbf{to} \,\, M$}
\Statex  $\quad \quad \quad${[}Simulate random samples {]}
\State $X^{m}_{n} := Rand \in \mathcal{X}$\Comment{State}
\State $\widetilde{\pi}^m_{n} := Rand \in \mathcal{A}_n$\Comment{Control}
\State $z^m_{n+1} := Rand \in \mathcal{Z}$\Comment{Disturbance}
\Statex $\quad \quad \quad${[}Compute the state variable after control{]}
\State $\widetilde{X}^{m}_{n+1} := \mathcal{T}_n(X^{m}_{n},\widetilde{\pi}^m_n,z^m_{n+1})$\Comment{Evolution of state}
\State if $n=0$ then $\widetilde{X}^{m}_{0}={X}^{m}_{0}$
\EndFor
\EndFor
\end{algorithmic}
\end{algorithm}

\begin{algorithm}[H]
\footnotesize
\caption{Backward solution (Realised Value Regress Now)}
\label{montecarlo_algorithm_realized}
\begin{algorithmic}[1]
\For{${n} = N \,\, \mathbf{to} \,\, 0$}
\If{$n = N$}  $\widehat{V}_{n}(\widetilde{{X}}_{n}) := R_N(\widetilde{{X}}_{n})$
\ElsIf{$n < N$}
\Statex $\quad \quad \quad${[}Perform regression of value function{]}
\State $\widehat{\bm{\theta}}_{n} := \arg \min_{\bm{\theta}_n} \sum^M_{m=1} \left[f_{\theta_n}(X^m_{n}, \widetilde{\pi}_{n}) -  \beta \widehat{V}_{n+1}(\widetilde{X}^m_{n+1}) \right]^2$
\Statex $\quad \quad \quad$Approximate conditional expectation $\widehat{\Phi}_{n}(X_{n},\widetilde{\pi}_{n})=f_{\widehat\theta_n}(X_{n}, \widetilde{\pi}_{n})$
\For{$m = 1 \,\, \mathbf{to} \,\, M$}
\State $\widehat{X}^m_{n} := \widetilde{X}^{m}_{n}$ \label{line:forwardsim}
\Statex $\quad \quad \quad \quad$ {[}Optimal control{]}
\State $\pi^*_{n}(\widehat{X}^{m}_{n}) := \arg \sup_{\pi_{n} \in A} \left\{ R_{n}(\widehat{X}^{m}_{n}, \pi_{n}) + \widehat{\Phi}_{n}(\widehat{X}^{m}_{n}, \pi_{n}) \right\}$ \label{line:optimalcontrol}
\Statex $\quad \quad \quad \quad$ {[}Update value function with optimal paths{]}
\State $\widehat{V}_{n}(\widehat{X}^{m}_{n}) := R_{n}(\widehat{X}^m_{n}, \pi^*_{n}(\widehat{X}^m_{n}))$
\State $\widehat{X}^m_{n+1} := \mathcal{T}_n(\widehat{X}^m_{n},\pi^*_{n}(\widehat{X}^m_{n}),z^m_{n})$
\For{$j = n + 1 \,\, \mathbf{to} \,\, N-1$} \label{line:loopforward}
\State $\widehat{V}_{n}(\widehat{X}^{m}_{n}) := \widehat{V}_{n}(\widehat{X}^{m}_{n}) + \beta_{n,j} R_{j}(\widehat{X}^m_{j}, \pi^*_{j}(\widehat{X}^m_{j}))$ \label{line:valuefunction}
\State $\widehat{X}^m_{j+1} := \mathcal{T}_j(\widehat{X}^m_{j},\pi^*_{j}(\widehat{X}^m_{j}),z^m_{j})$
\EndFor
\State $\widehat{V}_{n}(\widehat{X}^{m}_{n}) := \widehat{V}_{n}(\widehat{X}^{m}_{n}) + \beta_{n,N} R_N(\widehat{X}^m_{N})$
\EndFor
\EndIf
\EndFor
\end{algorithmic}
\end{algorithm}


\begin{algorithm}[H]
\footnotesize
\caption{Backward solution (Regression Surface Regress Now)}
\label{montecarlo_algorithm_regression}
\begin{algorithmic}[1]
\For{${n} = N \,\, \mathbf{to} \,\, 0$}
\If{$n = N$} $\widehat{V}_{n}(\widetilde{{X}}_{n}) := R_N(\widetilde{{X}}_{n})$
\ElsIf{$n < N$}
\Statex $\quad \quad \quad${[}Perform regression of transformed value function{]}
\State $\widehat{\bm{\theta}}_{n} := \arg \min_{\bm{\theta}_n} \sum^M_{m=1} \left[f_{\theta_n}(X^m_{n}, \widetilde{\pi}_{n}) -  \beta \widehat{V}_{n+1}(\widetilde{X}^m_{n+1}) \right]^2$
\Statex $\quad \quad \quad${[}Approximate conditional expectation $\widehat{\Phi}_{n}(X_{n},\widetilde{\pi}_{n})=f_{\widehat\theta_n}(X_{n}, \widetilde{\pi}_{n})$ {]}
\For{$m = 1 \,\, \mathbf{to} \,\, M$}
\State $\widehat{X}^m_{n} := \widetilde{X}^{m}_{n}$
\Statex $\quad \quad \quad \quad$ {[}Optimal control{]}
\State $\pi^*_{n}(\widehat{X}^{m}_{n}) := \arg \sup_{\pi_{n} \in A} \left\{ R_{n}( \widehat{X}^{m}_{n}, \pi_{n}) + \widehat{\Phi}_{n}(\widehat{X}^{m}_{n}, \pi_{n}) \right\}$
\State $\widehat{V}_{n}(\widehat{X}^{m}_{n}) := R_{n}( \widehat{X}^m_{n}, \pi^*_{n}(\widehat{X}^m_{n})) + \widehat{\Phi}_{n}(\widehat{X}^{m}_{n}, \pi^*_{n}(\widehat{X}^m_{n}))$ \label{line:valuefunction_regress_surf}
\EndFor
\EndIf
\EndFor
\end{algorithmic}
\end{algorithm}


\begin{algorithm}[H]
\footnotesize
\caption{Backward solution (Realised Value Regress Later)}
\label{montecarlo_algorithm_realized_later}
\begin{algorithmic}[1]
\For{${n} = N \,\, \mathbf{to} \,\, 0$}
\If{$n = N$} $\widehat{V}_{n}(\widetilde{{X}}_{n}) := R_N(\widetilde{{X}}_{n})$ \label{line:terminalvalue_later}
\ElsIf{$n < N$}
\Statex $\quad \quad \quad${[}Perform regression of value function{]}
\State $\widehat{\bm{\theta}}_{n+1}:=\arg\min_{\bm{\theta}_{n+1}}\sum_{m=1}^{M}\left[f_{\theta_{n+1}}(\widetilde{X}^m_{n+1})-\beta\widehat{V}_{n+1}(\widetilde{X}_{n+1}^{m})\right]^{2}$
\Statex $\quad \quad \quad$Approximate  $\widehat{\Phi}_{n}(X_{n},\widetilde{\pi}_{n})$ by computing $\E\left[f_{\widehat{\theta}_{n+1}}({X}^m_{n+1})\left|X_{n};\widetilde{\pi}_n\right.\right]$ 
\For{$m = 1 \,\, \mathbf{to} \,\, M$}
\State $\widehat{X}^m_{n} := \widetilde{X}^{m}_{n}$ \label{line:forwardsim_later}
\Statex $\quad \quad \quad \quad$ {[}Optimal control{]}
\State $\pi^*_{n}(\widehat{X}^{m}_{n}) := \arg \sup_{\pi_{n} \in \mathcal{A}_n} \left\{ R_{n}(\widehat{X}^{m}_{n}, \pi_{n}) + \widehat{\Phi}_{n}(\widehat{X}^{m}_{n}, \pi_{n}) \right\}$ \label{line:optimalcontrol_later}
\Statex $\quad \quad \quad \quad$ {[}Update value function with optimal paths{]}
\State $\widehat{V}_{n}(\widehat{X}^{m}_{n}) := R_{n}(\widehat{X}^m_{n}, \pi^*_{n}(\widehat{X}^m_{n}))$
\State $\widehat{X}^m_{n+1} := \mathcal{T}_n(\widehat{X}^m_{n},\pi^*_{n}(\widehat{X}^m_{n}),z^m_{n})$
\For{$j = n + 1 \,\, \mathbf{to} \,\, N-1$} \label{line:loopforward_later}
\State $\widehat{V}_{n}(\widehat{X}^{m}_{n}) := \widehat{V}_{n}(\widehat{X}^{m}_{n}) + \beta_{n,j} R_{j}(\widehat{X}^m_{j}, \pi^*_{j}(\widehat{X}^m_{j}))$ \label{line:valuefunction_later}
\State $\widehat{X}^m_{j+1} := \mathcal{T}_j(\widehat{X}^m_{j},\pi^*_{j}(\widehat{X}^m_{j}),z^m_{j})$
\EndFor
\State $\widehat{V}_{n}(\widehat{X}^{m}_{n}) := \widehat{V}_{n}(\widehat{X}^{m}_{n}) + \beta_{n,N} R_N(\widehat{X}^m_{N})$
\EndFor
\EndIf
\EndFor
\end{algorithmic}
\end{algorithm}
\begin{algorithm}[!h]
\footnotesize
\caption{Backward solution (Regression Surface Regress Later)}
\label{algorithm_RegressionSurfaceRegressLater}
\begin{algorithmic}[1]
\For{${n} = N \,\, \mathbf{to} \,\, 0$}
\If{$n = N$} $\widehat{V}_{n}(\widetilde{{X}}_{n}) := R_N(\widetilde{{X}}_{n})$
\ElsIf{$n < N$}
\Statex $\quad \quad \quad${[}Regression of transformed value function{]}
\State $\widehat{\bm{\theta}}_{n+1} := \arg \min_{\bm{\theta}_{n+1}} \sum^M_{m=1} \left[f_{\theta_{n+1}}(\widetilde{X}^m_{n+1}) -  \beta \widehat{V}_{n+1}(\widetilde{X}^m_{n+1}) \right]^2$
\Statex $\quad \quad \quad${[}Approximate  $\widehat{\Phi}_{n}(X_{n},\widetilde{\pi}_{n})$ by calculating $\E\left[f_{\widehat{\theta}_{n+1}}({X}^m_{n+1})\left|X_{n};\widetilde{\pi}_n\right.\right]$ {]} 
\For{$m = 1 \,\, \mathbf{to} \,\, M$}
\State $\widehat{X}^m_{n} := \widetilde{X}^{m}_{n}$
\Statex $\quad \quad \quad \quad$ {[}Optimal control{]}
\State $\pi^*_{n}(\widehat{X}^{m}_{n}) := \arg \sup_{\pi_{n} \in \mathcal{A}_n} \left\{ R_{n}( \widehat{X}^{m}_{n}, \pi_{n}) + \widehat{\Phi}_{n}(\widehat{X}^{m}_{n}, \pi_{n}) \right\}$
\State $\widehat{V}_{n}(\widehat{X}^{m}_{n}) := R_{n}( \widehat{X}^m_{n}, \pi^*_{n}(\widehat{X}^m_{n})) + \widehat{\Phi}_{n}(\widehat{X}^{m}_{n}, \pi^*_{n}(\widehat{X}^m_{n}))$ \label{line:valuefunction_regress_surf_later}
\EndFor
\EndIf
\EndFor
\end{algorithmic}
\end{algorithm}

~

\newpage

~

\subsection{Upper and lower estimators}
All LSMC methods presented above, Algorithms \ref{montecarlo_algorithm_realized}-\ref{algorithm_RegressionSurfaceRegressLater}, estimate the optimal policy $\widehat\pi_n^\ast(x)$, $n=1,...,N-1$. The contract value for a given policy is given by \eqref{GMWDB_general_eq}.
If we replace the optimal policy by its estimate and calculate the expectation by averaging $M$ contract payoffs over independent Monte Carlo trajectories then, by definition of supremum, the obtained contract value estimator will be a lower biased estimator
\begin{equation}\label{MC_lower_estimator_eq}
\widehat{V}_0^L\left({X}_0\right)=\frac{1}{M}\sum_{m=1}^M H_0({X^m},\hat{{\pi}}^\ast).
\end{equation}
\emph{Realised Value} Algorithms \ref{montecarlo_algorithm_realized} and \ref{montecarlo_algorithm_realized_later} already calculate this lower estimator if we set $X_0^m=X_0, m=1,...,M$ and find $\widehat{V}^L_0=\frac{1}{M}\sum_m \hat{V}_0(\hat{X}_0^m)$. In the case of \emph{Regression Surface} Algorithms \ref{montecarlo_algorithm_regression} and \ref{algorithm_RegressionSurfaceRegressLater}, formula \eqref{MC_lower_estimator_eq} should be implemented once the optimal policy has been estimated. This regression surface estimator $\widehat{V}_0(X_0)$ calculated by Algorithms \ref{montecarlo_algorithm_regression} and \ref{algorithm_RegressionSurfaceRegressLater} can be shown to be an approximate upper estimator $\hat{V}_0^U(X_0)$, by using Jensen's inequality to swap the order of the expectation and supremum operators iteratively, see \citet[page~223]{bouchard2011montecarlo}, \citet[Appendix~C]{aid2014probabilistic}.

\section{Numerical study\label{sec:numerical_results}}
In this section we present LSMC results for GMWB prices under the optimal policyholder withdrawal strategies, compared with some
established accurate finite-difference results obtained with a fine mesh and small time steps. We show results for \emph{Regress Now} LSMC methods only. It is possible to obtain results for \emph{Regress Later} LSMC methods for models considered in our study, but these methods cannot be generalised to more complicated models where numerical integration is required for conditional expectations in which more than two stochastic state variables are involved. Also, in the case of neural network regression, conditional expectations would have to be calculated numerically regardless of the model complexity. For the benefit of the readers, \ref{app:regress_later} presents closed-form formulas for the moments of the state variables up to order three in the case of the Vasicek stochastic interest rate model that can be used for Regress Later LSMC, but we do not pursue numerical results here. We mention that we tried Regress Later LSMC with polynomial regression for a simple model with constant interest rate, but we found that it did not perform as well as Regress Now methods.

\subsection{LSMC settings}
For LSMC calculations we use the following settings for polynomial and neural network regressions.

\noindent\textbf{Polynomial regression}
\begin{enumerate}
\item For forward simulation we just use a standard forward simulation in Algorithm~\ref{simulation_algorithm}, i.e. line 3 is replaced with $X_t^m=\widetilde{X}_t^m$ and $X_0=\widetilde{X}_0=x_0$.
\item We do not perform any transformation of value functions before regression. 
\item Basis functions $\mathbf{L}(.)$ for Algorithms \ref{montecarlo_algorithm_realized} and \ref{montecarlo_algorithm_regression}:\\
In cases of constant interest rate, polynomials up to the 3rd order of $W, A, \pi$ are used, i.e. every term in the polynomial function takes the form $W^{j_w}A^{j_a}\pi^{j_\pi}$, where $j_w,j_a,j_\pi$ are all non-negative integers and $j_w+j_a+j_\pi \leq 3$. Initially all possible combinations of $j_w,\;j_a,\;j_\pi$ were used, i.e. the complete 3rd order three-dimensional polynomial function was used. Later, after some numerical experiments the terms $W^2\pi$, $A^2\pi$ and $\pi^3$ were removed from the set, because they did not improve the accuracy of results. For cases with stochastic interest rate, the basis functions are described in Section \ref{NumericalResults_subsec2}.
\end{enumerate}

\noindent\textbf{Neural network regression settings}\\
The neural network uses a fully connected feed forward architecture with input dimension $D_{\text{in}} = 2$ (post-decision state variables) for constant interest rates and $D_{\text{in}} = 3$ for stochastic interest rates, three hidden layers each of width $H = 128$ with SiLU activation (Sigmoid Linear Unit, $SiLU(x)=x/(1+e^{-x})$), and output dimension $D_{\text{out}} = 1$. Training is performed for $2{,}000$ epochs using the Adam optimizer with learning rate $0.001$, weight decay $10^{-5}$, and cosine annealing schedule. Optimal withdrawals are computed by grid search with $1{,}000$ discretization points. 

\subsection{Model and VA rider specification}\label{model_sec}
For numerical illustration we consider the joint dynamics for the reference portfolio of assets $S(t)$, e.g. a mutual fund, underlying the contract and the stochastic interest rate $r(t)$, under the risk-neutral probability measure $\mathbb{Q}$, governed by
\begin{eqnarray}\label{referenceportfolio_eq}
\begin{split}
\frac{dS(t)}{S(t)}&=r(t) dt +\sigma_S \left(\rho d\mathcal{B}_1(t) +\sqrt{1-\rho^2} d\mathcal{B}_2(t)\right),\\
dr(t)&=\kappa(\theta-r(t))dt+\sigma_r d\mathcal{B}_1(t).
\end{split}
\end{eqnarray}
Here, $\mathcal{B}_1(t)$ and $\mathcal{B}_2(t)$ are independent standard Wiener processes, $\rho$ is the correlation coefficient between $S(t)$ and $r(t)$ processes, and $\sigma_S$ is the asset volatility parameter. The process for the interest rate $r(t)$ is the well-known Vasicek model with constant parameters $\kappa$, $\theta$ and $\sigma_r$. For simplicity of notation, we assume that model parameters are constant in time though the results can be generalized to the case of time dependent parameters. 

For this stochastic interest rate model, the price of a zero coupon bond $P(t,T)$ at time $t$ with maturity $T$, can be found in closed-form and then a change of num\'eraire can simplify the calculation of expectations; see \citet{shevchenko2017gmwb}. We will not pursue this in this paper but rather consider a brute force LSMC because this specific change of numeraire trick may not work for other interest rate models.


Consider the following VA contract with a basic GMWB often used in research studies, which is convenient for benchmarking. The actual products may have extra features, but these can be easily incorporated in the model and numerical algorithms developed in this paper.
\begin{itemize}
\item The premium paid by the policyholder upfront at $t_0$ is invested into the reference portfolio/risky asset $S(t)$.
The value of this portfolio (hereafter
referred to as \emph{\textbf{wealth account}}) at time $t$ is denoted as $W(t)$, so that
the upfront premium paid by the policyholder is $W(0)$. GMWB guarantees
the return of the premium via the withdrawals $\pi_n\ge 0$ allowed at
times $t_n$, $n=1,2,\ldots,N$.
The total amount of withdrawals cannot exceed the guarantee $W(0)$,
and withdrawals can be different from the contractual (guaranteed)
withdrawal $G_n=W(0)(t_n-t_{n-1})/T$, with penalties imposed if
$\pi_n>G_n$. Denote the annual contractual rate as $g:=1/T$.
Then, the wealth account $W(t)$ evolves as
\begin{equation}\label{eq_Wt}
W(t_{n+1})=\max\left(W(t_n)-\pi_n,0\right)\frac{S(t_{n+1})}{S(t_{n})} e^{-\alpha \Delta_{n+1}},\\
\end{equation}

where $\Delta_n=t_n-t_{n-1}$ and \emph{$\alpha$ is the annual fee continuously charged by the contract issuer}. If the account balance
becomes zero or negative, then it will stay zero till maturity.

\item Denote the value of the contract guarantee at time $t$ as $A(t)$, hereafter referred to as \emph{\textbf{guarantee account}}, with $A(0)=W(0)$. The guarantee balance
evolves as
\begin{equation}\label{accountbalance_eq}
A(t_{n+1})=A(t_n)-\pi_n\;\; n=1,2,\ldots,N-1
\end{equation}
with $\pi_n\le A(t_{n})$ 

\item The cashflow received by the policyholder at the withdrawal time $t_n$ is given by
\begin{equation}
C_n(\pi_n)=\left\{\begin{array}{ll}
                   \pi_n, & \mathrm{if}\; 0\le \pi_n\le G_n, \\
                   G_n+(1-\beta)(\pi_n-G_n), & \mathrm{if}\; \pi_n>G_n,
                 \end{array} \right.
\end{equation}
where $G_n$ is the contractual withdrawal and $\beta\in
[0,1]$ is the penalty coefficient applied to the portion of withdrawal above
$G_n$.

\item Let $V_n(W,r,A)$ be a price of the VA contract with GMWB at time $t_n$, when $W(t)=W$, $r(t)=r$, $A(t)=A$. At maturity, the policyholder takes the maximum between
the remaining guarantee account net
of penalty charge and the remaining balance of the wealth account,
i.e. the final payoff is
\begin{equation}\label{finalcond_eq}
V_{N}(W,r, A)=\max\left(W,C_N(A)\right).
\end{equation}

\end{itemize}

During the contract, the policyholder receives the cashflows $C_n(\pi_n)$,
$n=1,2,\ldots,N-1$ and the final payoff at maturity. 
Given the withdrawal strategy $\bm{\pi}=(\pi_1,\ldots,\pi_{N-1})$, the present
value of the total contract payoff is given by \eqref{total_payoff_eq} with settings $X_t=(W(t),r(t),A(t))$, $R_N(X_n):=\max\left(W,C_N(A)\right)$, and $R_n(X_n,\pi_n):=C_n(\pi_n)$.

\subsection{GMWB pricing results for constant interest rate}\label{NumericalResults_subsec1}
In this section, we consider the case of constant interest rate and GMWB with parameter settings: $S(0)=1.0$, $r = 5\%$, $g=10\%$ ($T:=1/g=10$ years), $\beta=10\%$,
$N_w=1$, $\alpha = 0.0135$.
All LSMC price results are averages over 20 independent LSMC runs, and the corresponding standard errors are given in brackets next to the price estimates.
Table \ref{tab_dynamic1} shows price estimates for a 10-year GMWB contract with annual withdraw frequency at different volatility levels in the case of optimal withdrawal strategy. The reported results are from LSMC \emph{Regress Now/Realised Value} Algorithm~\ref{montecarlo_algorithm_realized} and \emph{Regress Now/Regression Surface} Algorithm~\ref{montecarlo_algorithm_regression}, both in the case of polynomial OLS regression and neural network regression.

\begin{table}[!h]\captionsetup{width=0.85\textwidth}\caption{\footnotesize{Prices of VA with GMWB under dynamic withdrawal strategy for different volatilities. The number of trajectories used in LSMC is $M=1{,}000{,}000$.}} \label{tab_dynamic1}
\begin{center}
{\footnotesize{\begin{tabular*}{\textwidth}{llllll} \toprule
 $\sigma \%$ &   Finite Difference & OLS Algo 2, $V^L$ &  OLS Algo 3, $V^U$ &  NN Algo 2, $V^L$ & NN Algo 3, $V^U$ \\
 \midrule
\enskip 5   & 0.92660 (0.00113) & 0.92616 (0.00002) & 0.92602
(0.00003) & 0.92406 (0.00009) & 0.93320 (0.00023)\\
10  & 0.94463 (0.00050) & 0.94260 (0.00005) & 0.94532 (0.00010) & 0.94042 (0.00016) & 0.95061 (0.00014)\\
15  & 0.96991 (0.00024) & 0.96588 (0.00007) & 0.97052 (0.00019) & 0.96386 (0.00038) &0.97749 (0.00035)\\
20  & 0.99763 (0.00011) & 0.99121 (0.00016) & 0.99969 (0.00033) & 0.98967 (0.00034) &1.00586 (0.00048)\\
\bottomrule
\end{tabular*}
}}\end{center}
\end{table}


For the benchmark results in Table \ref{tab_dynamic1}, we use results from the finite difference method (column ``FD''), where we have used 800 node points for $W$, 200 node points for $A$, and the time step size $dt=0.0125$, which is the typical setting to get very accurate results, as shown, e.g., in \citet{luo2015fast}. 
Because we do not know the exact solution, FD results will be used as the ``benchmark'' solution with which all the other numerical results will be compared to estimate their accuracy. To estimate the accuracy of FD results, we calculated FD results when the number of nodes for both $W$ and $A$ is half of the fine mesh, i.e. 400 node points for $W$, 100 node points for $A$, and the time step is halved to $dt=0.025$. It can be quantitatively shown (as described in, e.g., \cite[Appendix]{luo2015pricing}) that in general the difference between FD results based on the coarse mesh and the results based on the finer mesh with doubled number of nodes, can be used as an accuracy measure of the FD results of the coarser mesh. Thus, the reported FD error is conservative as it is reported for the fine mesh in our table.



In all LSMC methods, the randomized distribution of withdraw amount in the forward simulation is a combination of discrete uniform and continuous uniform. Specifically, we give $25\%$ probability to each of no withdraw ($dW=0$) and withdraw at the contractual rate ($dW=G$), and $50\%$ probability to the withdraw amount $0 < dW < A(t)$ (uniform and continuous). This randomization gave the best results in our experiments.

As can be seen, comparing LSMC results with the FD results, the regression surface Algorithm~\ref{montecarlo_algorithm_regression} upper estimator $V^U$ obtained using a 3rd order polynomial regression performs a bit better than in the case of NN regression. Lower estimators from the NN and OLS regressions are very close. 



Table \ref{tab_dynamic2} shows LSMC results as the number of trajectories $M$ increases from $10^4$ to $4\times 10^6$ indicating that one needs $M>10^5$ trajectories to get at least 1\% accuracy. It also shows that realised value LSMC works better when the number of trajectories is small, and is outperformed by regression surface LSMC when $M\ge 10^6$. We also note that in this table we show results both for $V^L$ and $V^U$ estimators from Algorithm~\ref{montecarlo_algorithm_regression}. Comparing to regression surface, $V^L$ for realised value performs better for small $M$ and the same for large $M$ and thus in all other tables we show $V^L$ from realised value LSMC Algorithm~\ref{montecarlo_algorithm_realized} only. 

Overall, in the reported and other results, we observed that $V^L$ is lower and $V^U$ is larger than the ``exact'' result, and these bounds converge towards the ``exact'' value as $M$ increases. As expected, the standard errors of the estimators decrease as $M$ increases.
Also, as expected, the value of the contract and its standard error increase as volatility $\sigma$ increases.

We also tried post decision variables \eqref{equation-post-decision}  OLS regressions with polynomials up to order 3, but the obtained accuracy was not as good as for OLS regressions reported in Table \ref{tab_dynamic1}. This is another highlight of the NN advantage that it does not require such feature engineering.

Finally we note that the computing time in the case of the polynomial regression results reported in Table \ref{tab_dynamic1} is approximately 40 minutes over 20 runs (i.e. about 2 minutes per price calculation) on a modern laptop (Intel(R) Core(TM) Ultra 7 255U, 2.00 GHz) without parallel threading using Intel Fortran. Of course, the actual total computational effort is substantially larger, as it also involves experimentation with different basis functions. The computing time in the case of the neural network regression results reported in Table \ref{tab_dynamic1} is approximately 3 hours over 20 runs (i.e. about 9 minutes per price calculation) using PyTorch with GPU acceleration running on an NVIDIA H20 GPU.

\begin{table}[!h]\captionsetup{width=0.85\textwidth}\caption{\footnotesize{Regress Now algorithms in the case of polynomial regression, constant interest rate for $\sigma=0.2$. Other parameters and settings are the same as for Table \ref{tab_dynamic1}.}} \label{tab_dynamic2}
\begin{center}
{\footnotesize{\begin{tabular*}{0.65\textwidth}{llll} \toprule
 $M$ &   OLS Algo 2, $V^L$ &  OLS Algo 3, $V^L$ &  OLS Algo 3, $V^U$  \\
 \midrule
 $10^4$ &  0.97372 (0.00130)&   0.93860 (0.00287)      &1.23084 (0.03320)\\ 
 $10^5$ & 0.98721 (0.00057) &0.96850 (0.00423)&1.01962 (0.00413)\\
$10^6$   & 0.99121 (0.00016) &  0.99139 (0.00026) & 0.99969 (0.00033) \\
$2\!\times\!10^6$  & 0.99161 (0.00010) &  0.99212 (0.00022) & 0.99882 (0.00033) \\
$4\!\times\!10^6$  & 0.99164 (0.00007) &  0.99229 (0.00010) & 0.99817 (0.00012) \\
\bottomrule
\end{tabular*}
}}\end{center}
\end{table}


\subsection{GMWB pricing results for stochastic interest rate}\label{NumericalResults_subsec2}
In this section we show GMWB pricing results in the case of the Vasicek stochastic interest rate model. The simulation of the underlying model~\eqref{referenceportfolio_eq} is done using the exact simulation scheme described in \ref{app:exact_simulation}. Results obtained by LSMC are compared with those obtained by GHQC (Gauss-Hermite quadrature on
cubic spline) developed in \citet{luo2014fast} and used in
\citet{shevchenko2017gmwb} for pricing GMWB in the case of stochastic interest rate\footnote{We use the same code for GHQC method as described and used in
\citet{shevchenko2017gmwb}.}. Unlike the previous section where some very fine finite-difference results were used as a benchmark, here we do not have high accuracy finite-difference results with very fine meshes due to the higher dimension of the problem.

In Table \ref{tab_vasicek1}, LSMC pricing results (column 3 to column 6) for a 10-year GMWB contract with annual withdraw frequency for different values of volatility $\sigma$ under optimal dynamic withdrawals are shown for the OLS and NN regressions, in comparison with GHQC results (column 2).
For LSMC, we show numerical results for price estimates (obtained as an average over 100 independent runs of the LSMC calculations) and their corresponding standard errors.
All the input parameters are given in the table caption. The numerical setup (e.g. number of quadrature points) for GHQC is exactly the same as in \citet{shevchenko2017gmwb}. For OLS LSMC calculations we used the complete set of 2nd order polynomial basis functions, that is, all the terms in the form of $W^{j_w}A^{j_a}\pi^{j_\pi} r^{j_r}$ are included, where $j_w,j_a,j_\pi,j_r$ are all non-negative integers and $j_w+j_a+j_\pi +j_r \leq 2$, totalling 15 terms altogether including the constant term. 

\begin{table}[!h]\captionsetup{width=0.87\textwidth}\caption{\footnotesize{Prices of VA with GMWB under dynamic withdrawal strategy in the case of Vasicek interest rate model for different volatilities. Input parameters: $S(0)=1.0$, $r(0) =\theta= 5\%$, $\kappa=0.0349$,$\sigma_r=2\%$, $\rho=0.3$, $g=10\%$ ($T=1/g$), $\beta=10\%$,
$N_w=1$, $\alpha = 0.01$. LSMC results are the price average and corresponding standard error (in brackets next to the price estimate) over 100 independent LSMC runs. The number of trajectories is $M=1{,}000{,}000$.}} \label{tab_vasicek1}
\begin{center}
\hspace{-.5cm}{\footnotesize{\begin{tabular*}{0.9\textwidth}{llllll} \toprule
 $\sigma \%$ &   GHQC &  OLS Algo 2, $V^L$  &  OLS Algo 3, $V^U$ &  NN Algo 2, $V^L$ &NN Algo 3, $V^U$ \\
 \midrule
5   & 0.95325 & 0.95269 (0.00532) &0.95560 (0.00534)   &  0.95395 (0.00013) & 0.95641 (0.00005)
\\
10  & 0.97411 & 0.96976 (0.00542) & 0.97495 (0.00511)   &  0.97191 (0.00015) & 0.98215 (0.00009)
\\
15  & 1.00147 & 0.99191 (0.00554)&1.00500 (0.00559)   &  0.99814 (0.00023) & 1.02284 (0.00086)
\\
20  & 1.03436 & 1.01692 (0.00568) & 1.03888 (0.00547)   &  1.03748 (0.00022) & 1.07160 (0.00134)
\\
\bottomrule
\end{tabular*}
}}
\end{center}
\end{table}

We also tried other sets of polynomial basis functions. For example, we considered the set with terms in the form $W^{j_w}A^{j_a}\pi^{j_\pi}$ (without containing interest rate $r$), the same as in the case of constant interest rate used in the previous section (e.g. Table \ref{tab_dynamic1}), and all the extra terms involving $r$ have the form $W^{j_w}A^{j_a}\pi^{j_\pi} r^{j_r}$ with $1\leq j_w+j_a+j_\pi +j_r \leq 2$ and $1\leq j_r \leq 2$. In other words, the extra terms involving $r$ are $rW$, $rA$,$r\pi$, $r$ and $r^2$. Altogether in this set there were 23 terms. Results obtained with this set are not materially different from the reported ones (i.e. the difference is well within standard errors). We also tried to add all the cubic terms involving $r$ to set B, i.e. adding all possible terms of the form $W^{j_w}A^{j_a}\pi^{j_\pi} r^{j_r}$, with the constraints $1\leq j_r \leq 3$ and $ j_w+j_a+j_\pi +j_r =3$. This set has altogether 30 polynomial terms. Again, we did not observe material difference in results.

Taking the GHQC results as fairly accurate with relative error of the order of $10^{-4}$ (see \citet{luo2015gmwd} and \citet{shevchenko2017gmwb}), we can measure the performance of LSMC.
From Table \ref{tab_vasicek1}, we can see that the OLS Algorithm~\ref{montecarlo_algorithm_regression} upper estimator $V^U$ appears above the `true' value and the OLS Algorithm~\ref{montecarlo_algorithm_realized} lower estimator $V^L$ appears below the `true' value.
The upper and lower NN LSMC estimates also appear to be consistent with `true' values, though OLS estimators appear slightly more accurate. 
Also, we can see that the standard errors of OLS LSMC estimators are now one order of magnitude larger than in the cases with constant interest rate shown in the previous section. By contrast, the standard errors of NN estimators are similar to the constant interest rate case. The main reason for this phenomenon is that the neural network architecture scales gracefully with the dimensionality of the state space: extending the model from the constant interest rate case ($D_{\text{in}} = 2$) to the stochastic interest rate case ($D_{\text{in}} = 3$) increases the total number of parameters only marginally, from $33{,}537$ to $33{,}665$ (an increase of $128$ parameters, or about $0.4\%$), since the additional input dimension only affects the weight matrix of the first hidden layer ($D_{\text{in}} \times H$). This is in sharp contrast to polynomial regression, where the number of basis functions grows combinatorially with the number of state variables, as illustrated above by the expansion from the original basis set to the 15-term set when including interest rates. The NN approach therefore generalizes to higher-dimensional state spaces with essentially no additional model complexity, leading to similar standard errors, and also without the need for manual feature engineering.

\section{Conclusion\label{sec:conclusion}}
In this paper, we demonstrated that the numerical evaluation of VAs under optimal withdrawal strategies can be successfully accomplished using the LSMC method.
The main advantages of the method are its flexibility with respect to the underlying stochastic processes, its straightforward extension to additional state and control variables, and its ability to handle complex contract payoff structures. In our experiments, polynomial regression performed slightly better than NN regression. However, polynomial regression required substantially more preliminary analysis to identify suitable basis functions. 
Overall, we recommend the use of NN regression because, due to its ``non-parametric'' nature, it can adapt to the shape of the target function without requiring the functional form to be specified in advance. A similar observation was made in macroeconomic models studied by \citet{valaitis2024machine} where NN regression was advocated over polynomial regression for approximating conditional expectations in macro-finance models. Moreover, making interest rates stochastic did not cause any trouble to the deep LSMC method, while this caused the standard errors of classical polynomial LSMC to increase by one order of magnitude.

We presented and discussed different variants of LSMC: regress now/regress later, regression surface/realised value. It is important to note that LSMC estimators are downward or upward biased depending on the method, and it is prudent to report both the upper and lower estimators. The difference between the upper and lower estimators is significantly larger than standard errors of the estimators obtained over independent runs.

Regression-based approximations of conditional expectations have long been used in computational 
economics through forward-in-time simulation combined with regression to evaluate expectations arising in equilibrium conditions \citep{den1990solving}. This approach was later extended through the use of neural networks by \cite{duffy2001approximating}, while \citet{valaitis2024machine} revisited related ideas from a modern machine-learning perspective in macro-finance models. In the case of LSMC, we solve a \emph{recursive stochastic control} problem backward in time by approximating the conditional expectation in the Bellman equation via regression, and then recover the optimal control though a separate numerical maximization. This separation is particularly convenient for VAs because the controls are constrained and the state dynamics is endogenous.

Finally, we mention that we attempted to price GMWB using a global neural network parameterisation of controls with time as an additional input variable and then estimating the NN parameters by maximizing the empirical objective function directly by gradient ascent. While easier to implement, this approach did not reach the same accuracy as LSMC. We therefore leave the exploration of this approach for further research.

\section*{Acknowledgement}
Nicolas Langren\'e acknowledges the support of the Guangdong Provincial/Zhuhai Key Laboratory of IRADS (2022B1212010006).

\bibliographystyle{plainnat_lastnamefirst} 
\bibliography{bibliography}

\appendix

\renewcommand{\baselinestretch}{1.0}

\section{Closed-form formulas}\label{appendix1}

We consider the model
\begin{align}
dS(t) & =r(t)S(t)dt+\sigma_{S}S(t)(\rho d\mathcal{B}_{1}(t)+\sqrt{1-\rho^{2}}d\mathcal{B}_{2}(t))\label{eq:dSt}\\
dr(t) & =\kappa(\theta-r(t))dt+\sigma_{r}d\mathcal{B}_{1}(t)\label{eq:drt}
\end{align}

\subsection{Exact simulation scheme\label{app:exact_simulation}}

The strong solutions of $S(t)$ and $r(t)$ are given by:
\begin{align*}
S(t) & =S(0)\exp\left(\int^{t}_{0}\left(r(\tau)-\frac{1}{2}\sigma^{2}_{S}\right)d\tau+\int^{t}_{0}\sigma_{S}\left(\rho d\mathcal{B}_{1}(\tau)+\sqrt{1-\rho^{2}}d\mathcal{B}_{2}(\tau)\right)\right)\\
r(t) & =e^{-\kappa t}\left(r(0)+\int^{t}_{0}e^{\kappa\tau}\kappa\theta d\tau+\int^{t}_{0}e^{\kappa\tau}\sigma_{r}d\mathcal{B}_{1}(\tau)\right)
\end{align*}
Similarly, between times $t_{i}$ and $t_{i+1}$ with $0\leq t_{i}<t_{i+1}$:
\begin{align*}
S(t_{i+1}) & =S(t_{i})\exp\left(\int^{t_{i+1}}_{t_{i}}r(\tau)d\tau-\frac{1}{2}\sigma^{2}_{S}\Delta_{t_{i}}+\int^{t_{i+1}}_{t_{i}}\sigma_{S}\left(\rho d\mathcal{B}_{1}(\tau)+\sqrt{1-\rho^{2}}d\mathcal{B}_{2}(\tau)\right)\right)\\
r(t_{i+1}) & =e^{-\kappa\Delta_{t_{i}}}\left(r(t_{i})+\int^{t_{i+1}}_{t_{i}}e^{\kappa(\tau-t_{i})}\kappa\theta d\tau+\int^{t_{i+1}}_{t_{i}}e^{\kappa(\tau-t_{i})}\sigma_{r}d\mathcal{B}_{1}(\tau)\right)
\end{align*}
where $\Delta_{t_{i}}:=t_{i+1}-t_{i}$. Introduce the following three
random variables:
\begin{align}
R:= & r(t_{i+1})-e^{-\kappa\Delta_{t_{i}}}r(t_{i})=e^{-\kappa\Delta_{t_{i}}}\left(\int^{t_{i+1}}_{t_{i}}e^{\kappa(\tau-t_{i})}\kappa\theta d\tau+\int^{t_{i+1}}_{t_{i}}e^{\kappa(\tau-t_{i})}\sigma_{r}d\mathcal{B}_{1}(\tau)\right)\label{eq:R}\\
Y:= & \int^{t_{i+1}}_{t_{i}}\!(r(\tau)-e^{-\kappa\Delta_{t_{i}}}r(t_{i}))d\tau=\int^{t_{i+1}}_{t_{i}}\!\!e^{-\kappa(u-t_{i})}\left(\int^{u}_{t_{i}}\!e^{\kappa(\tau-t_{i})}\kappa\theta d\tau+\!\int^{u}_{t_{i}}\!e^{\kappa(\tau-t_{i})}\sigma_{r}d\mathcal{B}_{1}(\tau)\right)du\label{eq:Y}\\
E:= & -\frac{1}{2}\sigma^{2}_{S}\Delta_{t_{i}}+\int^{t_{i+1}}_{t_{i}}\sigma_{S}\left(\rho d\mathcal{B}_{1}(\tau)+\sqrt{1-\rho^{2}}d\mathcal{B}_{2}(\tau)\right)\label{eq:E}
\end{align}
These three random variables are normally distributed and are independent
of $\mathcal{F}_{t_{i}}$. Using these notations, we can write $S(t_{i+1})$
and $r(t_{i+1})$ as
\begin{align}
S(t_{i+1}) & =S(t_{i})\exp\left(e^{-\kappa\Delta_{t_{i}}}r(t_{i})\Delta_{t_{i}}+Y+E\right)\label{eq:S_stepping}\\
r(t_{i+1}) & =e^{-\kappa\Delta_{t_{i}}}r(t_{i})+R\label{eq:r_stepping}
\end{align}
where $\Delta_{t_{i}}:=t_{i+1}-t_{i}$. To complete the definition
of the exact simulation scheme \eqref{eq:S_stepping}-\eqref{eq:r_stepping},
we need to compute the mean, variance and correlations of the three
Gaussian random variables $R$, $Y$ and $E$. We will make use of
the following classical lemma.
\begin{lem}
\label{lem:gaussian_decomposition}If $N_{1}$ and $N_{2}$ are two
standard Gaussian random variables with correlation $\rho_{12}$,
then
\begin{equation}
N_{1}=\rho_{12}N_{2}+\sqrt{1-\rho^{2}_{12}}N^{\perp}_{2}\label{eq:2N_decomposition}
\end{equation}
where $N^{\perp}_{2}$ is a standard Gaussian random variable independent
of $N_{2}$.
\end{lem}
We can now compute the mean and variance of $R$, $Y$ and $E$. The
following formulas hold:
\begin{align}
R & =e^{-\kappa\Delta_{t_{i}}}\left(\int^{t_{i+1}}_{t_{i}}e^{\kappa(\tau-t_{i})}\kappa\theta d\tau+\int^{t_{i+1}}_{t_{i}}e^{\kappa(\tau-t_{i})}\sigma_{r}d\mathcal{B}_{1}(\tau)\right)=\mu_{R}+\varsigma_{R}N_{R}\label{eq:R_decomposition}\\
\mu_{R} & :=\E[R]=e^{-\kappa\Delta_{t_{i}}}\int^{t_{i+1}}_{t_{i}}e^{\kappa(\tau-t_{i})}\kappa\theta d\tau=\theta\left(1-e^{-\kappa\Delta_{t_{i}}}\right)\nonumber \\
\varsigma^{2}_{R} & :=\Var[R]=\frac{\sigma^{2}_{r}}{2\kappa}\left[1-e^{-2\kappa\Delta_{t_{i}}}\right]\nonumber \\
N_{R} & :=\varsigma^{-1}_{R}e^{-\kappa\Delta_{t_{i}}}\int^{t_{i+1}}_{t_{i}}e^{\kappa(\tau-t_{i})}\sigma_{r}d\mathcal{B}_{1}(\tau)\ \sim\ \mathcal{N}(0,1)\nonumber 
\end{align}

\begin{align}
Y & =\int^{t_{i+1}}_{t_{i}}e^{-\kappa(u-t_{i})}\left(\int^{u}_{t_{i}}e^{\kappa(\tau-t_{i})}\kappa\theta d\tau+\int^{u}_{t_{i}}e^{\kappa(\tau-t_{i})}\sigma_{r}d\mathcal{B}_{1}(\tau)\right)du=\mu_{Y}+\varsigma_{Y}N_{Y}\label{eq:Y_decomposition}\\
\mu_{Y} & :=\E[Y]=\int^{t_{i+1}}_{t_{i}}e^{-\kappa(u-t_{i})}\left(\int^{u}_{t_{i}}e^{\kappa(\tau-t_{i})}\kappa\theta d\tau\right)du\nonumber \\
 & =\theta\left(\Delta_{t_{i}}-\frac{1-e^{-\kappa\Delta_{t_{i}}}}{\kappa}\right)\nonumber \\
\varsigma^{2}_{Y} & :=\Var[Y]=\frac{\sigma^{2}_{r}}{\kappa^{2}}\int^{t_{i+1}}_{t_{i}}\left(1-e^{-\kappa(\tau-t_{i})}\right)^{2}d\tau\nonumber \\
 & =\frac{\sigma^{2}_{r}}{\kappa^{2}}\left[\Delta_{t_{i}}-\frac{2}{\kappa}\left(1-e^{-\kappa\Delta_{t_{i}}}\right)+\frac{1}{2\kappa}\left(1-e^{-2\kappa\Delta_{t_{i}}}\right)\right]\nonumber \\
N_{Y} & :=\varsigma^{-1}_{Y}\int^{t_{i+1}}_{t_{i}}e^{-\kappa(u-t_{i})}\int^{u}_{t_{i}}e^{\kappa(\tau-t_{i})}\sigma_{r}d\mathcal{B}_{1}(\tau)du\ \sim\ \mathcal{N}(0,1)\nonumber 
\end{align}

\begin{align}
E & =-\frac{1}{2}\sigma^{2}_{S}\Delta_{t_{i}}+\int^{t_{i+1}}_{t_{i}}\sigma_{S}\left(\rho d\mathcal{B}_{1}(\tau)+\sqrt{1-\rho^{2}}d\mathcal{B}_{2}(\tau)\right)=\mu_{E}+\varsigma_{E}N_{E}\label{eq:E_decomposition}\\
\mu_{E} & :=\E[E]=-\frac{1}{2}\sigma^{2}_{S}\Delta_{t_{i}}\nonumber \\
\varsigma^{2}_{E} & :=\Var[E]=\sigma^{2}_{S}\Delta_{t_{i}}\nonumber \\
N_{E} & =\varsigma^{-1}_{E}\int^{t_{i+1}}_{t_{i}}\sigma_{S}\left(\rho d\mathcal{B}_{1}(\tau)+\sqrt{1-\rho^{2}}d\mathcal{B}_{2}(\tau)\right)\ \sim\ \mathcal{N}(0,1)\nonumber 
\end{align}
So far, we established these three Gaussian decompositions \eqref{eq:R_decomposition}-\eqref{eq:Y_decomposition}-\eqref{eq:E_decomposition}:
\begin{align}
R & =\mu_{R}+\varsigma_{R}N_{R}\label{eq:R_Gaussian}\\
Y & =\mu_{Y}+\varsigma_{Y}N_{Y}\label{eq:Y_Gaussian}\\
E & =\mu_{E}+\varsigma_{E}N_{E}\label{eq:E_Gaussian}
\end{align}
The correlations between the three standard Gaussian random variables
$N_{R}$, $N_{Y}$ and $N_{E}$ are given by:
\begin{align}
\rho_{RY}:=\E[N_{R}N_{Y}] & =\frac{\sigma^{2}_{r}}{\varsigma_{r}\varsigma_{Y}}\frac{\left(1-e^{-\kappa\Delta_{t_{i}}}\right)^{2}}{2\kappa^{2}}\label{eq:rho_RY}\\
\rho_{RE}:=\E[N_{R}N_{E}] & =\frac{\rho\sigma_{r}\sigma_{S}}{\kappa\varsigma_{r}\varsigma_{E}}\left(1-e^{-\kappa\Delta_{t_{i}}}\right)\label{eq:rho_RE}\\
\rho_{YE}:=\E[N_{Y}N_{E}] & =\frac{\rho\sigma_{r}\sigma_{S}}{\kappa\varsigma_{Y}\varsigma_{E}}\left[\Delta_{t_{i}}-\frac{1}{\kappa}\left(1-e^{-\kappa\Delta_{t_{i}}}\right)\right]\label{eq:rho_YE}
\end{align}
To sum up, the model \eqref{eq:dSt}-\eqref{eq:drt} can be simulated
exactly on a discrete-time grid $0=t_{0}\leq t_{1}<\ldots<t_{N}=T$
by 
\begin{align}
S(t_{i+1}) & =S(t_{i})\exp\left(e^{-\kappa\Delta_{t_{i}}}r(t_{i})\Delta_{t_{i}}+\mu_{Y}+\mu_{E}+\varsigma_{Y}N_{Y}+\varsigma_{E}N_{E}\right)\label{eq:S_stepping_exact_1}\\
r(t_{i+1}) & =e^{-\kappa\Delta_{t_{i}}}r(t_{i})+\mu_{R}+\varsigma_{R}N_{R}\label{eq:r_stepping_exact_1}
\end{align}
where the standard Gaussian random variables $N_{Y}$, $N_{E}$ and
$N_{R}$ are correlated according to the formulas \eqref{eq:rho_RY}-\eqref{eq:rho_RE}-\eqref{eq:rho_YE}.
This can be further simplified into
\begin{align}
S(t_{i+1}) & =S(t_{i})\exp\left(e^{-\kappa\Delta_{t_{i}}}r(t_{i})\Delta_{t_{i}}+\mu_{Y}+\mu_{E}+\sqrt{\varsigma^{2}_{Y}+\varsigma^{2}_{E}+2\varsigma_{Y}\varsigma_{E}\rho_{YE}}N_{S}\right)\label{eq:S_stepping_exact_2}\\
r(t_{i+1}) & =e^{-\kappa\Delta_{t_{i}}}r(t_{i})+\mu_{R}+\varsigma_{R}N_{R}\label{eq:r_stepping_exact_2}\\
\rho_{RS} & =\E[N_{R}N_{S}]=\frac{\varsigma_{Y}\rho_{RY}+\varsigma_{E}\rho_{RE}}{\sqrt{\varsigma^{2}_{Y}+\varsigma^{2}_{E}+2\varsigma_{Y}\varsigma_{E}\rho_{YE}}}\label{eq:rho_RS}
\end{align}
where $N_{R}$ and $N_{S}$ are two standard Gaussian random variables
with correlation $\rho_{RS}$.

\subsection{Regress-later formulas\label{app:regress_later}}

\subsubsection{Formulas for general time step}

For the Regress-Later implementation with third-order polynomial basis,
we need to compute the discounted conditional moments $\E[\exp\left(-\int^{t_{i+1}}_{t_{i}}r(\tau)d\tau\right)S(t_{i+1})^{p}r(t_{i+1})^{q}\left|\mathcal{F}_{t_{i}}\right.]$
with $0\leq p\leq3$, $0\leq q\leq3$, $0<p+q\leq3$. We introduce
$\ell\in\mathbb{R}$ and consider the slightly more general problem
of computing
\begin{equation}
B_{\ell,p,q}:=\E[\exp\left(-\ell\int^{t_{i+1}}_{t_{i}}r(\tau)d\tau\right)S(t_{i+1})^{p}r(t_{i+1})^{q}\left|\mathcal{F}_{t_{i}}\right.]\label{eq:B}
\end{equation}
where we are interested in the case $\ell=1$, but the computations
easily accommodate any other value of $\ell$, including $\ell=0$
(no discounting). Using equations \eqref{eq:R}-\eqref{eq:Y}-\eqref{eq:E},
$B_{\ell,p,q}$ can be written as
\begin{equation}
B_{\ell,p,q}=S(t_{i})^{p}\exp\left((p-\ell)e^{-\kappa\Delta_{t_{i}}}r(t_{i})\Delta_{t_{i}}\right)\E\left[\exp\left((p-\ell)Y+pE\right)\left(e^{-\kappa\Delta_{t_{i}}}r(t_{i})+R\right)^{q}\left|\mathcal{F}_{t_{i}}\right.\right]\label{eq:B_decomposition}
\end{equation}
To compute the remaining conditional expectation in equation \eqref{eq:B_decomposition},
we need to use the formulas \eqref{eq:R_Gaussian}, \eqref{eq:Y_Gaussian}, \eqref{eq:E_Gaussian}, \eqref{eq:rho_RY}, \eqref{eq:rho_RE}, and \eqref{eq:rho_YE}
for $R$, $Y$, and $E$, as well as the following lemma:
\begin{lem}
\label{lem:Ecn}Let $N\sim\mathcal{N}(0,1)$ be a standard Gaussian
random variable, and let $c\in\mathbb{R}$ be a constant. Define
\begin{equation}
\mathcal{E}(c,n):=\E\left[N^{n}\exp\left(cN\right)\right]\label{eq:Ecn}
\end{equation}
The first few values of $\mathcal{E}(c,n)$ are given by 
\begin{align}
\mathcal{E}(c,0) & =\E\left[\exp\left(cN\right)\right]=e^{\frac{c^{2}}{2}}\label{eq:EecN}\\
\mathcal{E}(c,1) & =\E\left[N\exp\left(cN\right)\right]=ce^{\frac{c^{2}}{2}}\label{eq:ENecN}\\
\mathcal{E}(c,2) & =\E\left[N^{2}\exp\left(cN\right)\right]=(c^{2}+1)e^{\frac{c^{2}}{2}}\label{eq:EN2ecN}\\
\mathcal{E}(c,3) & =\E\left[N^{3}\exp\left(cN\right)\right]=(c^{3}+3c)e^{\frac{c^{2}}{2}}\label{eq:EN3ecN}
\end{align}
Subsequent values of $\mathcal{E}(c,n)$ can be obtained by the following
induction formula:
\[
\mathcal{E}(c,n)=(p-1)\mathcal{E}(c,n-2)+c\mathcal{E}(c,n-1)
\]
\end{lem}
\begin{lem}
We can now rewrite equation \eqref{eq:B_decomposition} as 
\begin{align}
B_{\ell,p,q} & =S(t_{i})^{p}\exp\left((p-\ell)e^{-\kappa\Delta_{t_{i}}}r(t_{i})\Delta_{t_{i}}\right)\nonumber \\
 & \times\E\left[\exp\left((p-\ell)(\mu_{Y}+\varsigma_{Y}N_{Y})+p(\mu_{E}+\varsigma_{E}N_{E})\right)\left(e^{-\kappa\Delta_{t_{i}}}r(t_{i})+\mu_{R}+\varsigma_{R}N_{R}\right)^{q}\left|\mathcal{F}_{t_{i}}\right.\right]\nonumber \\
 & =S(t_{i})^{p}\exp\left((p-\ell)(\mu_{Y}+e^{-\kappa\Delta_{t_{i}}}r(t_{i})\Delta_{t_{i}})+p\mu_{E}\right)\nonumber \\
 & \times\E\left[\exp\left((p-\ell)\varsigma_{Y}N_{Y}+p\varsigma_{E}N_{E}\right)\left(e^{-\kappa\Delta_{t_{i}}}r(t_{i})+\mu_{R}+\varsigma_{R}N_{R}\right)^{q}\left|\mathcal{F}_{t_{i}}\right.\right]\label{eq:B_decomposition_2}
\end{align}
Define 
\begin{equation}
C_{\ell,p,q}:=S(t_{i})^{p}\exp\left((p-\ell)(\mu_{Y}+e^{-\kappa\Delta_{t_{i}}}r(t_{i})\Delta_{t_{i}})+p\mu_{E}\right)\E\left[\exp\left((p-\ell)\varsigma_{Y}N_{Y}+p\varsigma_{E}N_{E}\right)N^{q}_{R}\right]\label{eq:Clpq}
\end{equation}
We have that
\begin{align}
B_{\ell,p,0} & =C_{\ell,p,0}\label{eq:BClp0}\\
B_{\ell,p,1} & =\left(e^{-\kappa\Delta_{t_{i}}}r(t_{i})+\mu_{R}\right)C_{\ell,p,0}+\varsigma_{R}C_{\ell,p,1}\label{eq:BClp1}\\
B_{\ell,p,2} & =\left(e^{-\kappa\Delta_{t_{i}}}r(t_{i})+\mu_{R}\right)^{2}C_{\ell,p,0}+2\left(e^{-\kappa\Delta_{t_{i}}}r(t_{i})+\mu_{R}\right)\varsigma_{R}C_{\ell,p,1}+\varsigma^{2}_{R}C_{\ell,p,2}\label{eq:BClp2}\\
B_{\ell,p,3} & =\left(e^{-\kappa\Delta_{t_{i}}}r(t_{i})+\mu_{R}\right)^{3}C_{\ell,p,0}+3\left(e^{-\kappa\Delta_{t_{i}}}r(t_{i})+\mu_{R}\right)^{2}\varsigma_{R}C_{\ell,p,1}\nonumber \\
 & +3\left(e^{-\kappa\Delta_{t_{i}}}r(t_{i})+\mu_{R}\right)\varsigma^{2}_{R}C_{\ell,p,2}+\varsigma^{3}_{R}C_{\ell,p,3}\label{eq:BClp3}
\end{align}
and so on, by expanding the power term $\left(e^{-\kappa\Delta_{t_{i}}}r(t_{i})+\mu_{R}+\varsigma_{R}N_{R}\right)^{q}$
in equation \eqref{eq:Clpq}. Remark that $(p-\ell)\varsigma_{Y}N_{Y}+p\varsigma_{E}N_{E}$
is a Gaussian random variable with mean and variance given by
\begin{align*}
\E\left[(p-\ell)\varsigma_{Y}N_{Y}+p\varsigma_{E}N_{E}\right] & =0\\
\Var[(p-\ell)\varsigma_{Y}N_{Y}+p\varsigma_{E}N_{E}] & =(p-\ell)^{2}\varsigma^{2}_{Y}+p^{2}\varsigma^{2}_{E}+2p(p-\ell)\rho_{YE}\varsigma_{Y}\varsigma_{E}
\end{align*}
Therefore
\begin{equation}
(p-\ell)\varsigma_{Y}N_{Y}+p\varsigma_{E}N_{E}=\sqrt{(p-\ell)^{2}\varsigma^{2}_{Y}+p^{2}\varsigma^{2}_{E}+2p(p-\ell)\rho_{YE}\varsigma_{Y}\varsigma_{E}}N_{YE}\label{eq:NY+NE}
\end{equation}
where $N_{YE}:=\frac{(p-\ell)\varsigma_{Y}N_{Y}+p\varsigma_{E}N_{E}}{\sqrt{(p-\ell)^{2}\varsigma^{2}_{Y}+p^{2}\varsigma^{2}_{E}+2p(p-\ell)\rho_{YE}\varsigma_{Y}\varsigma_{E}}}$
is a standard Gaussian variable, whose correlation with $N_{R}$ is
given by
\begin{align}
\rho_{RYE}:=\E[N_{R}N_{YE}] & =\frac{(p-\ell)\varsigma_{Y}\rho_{RY}+p\varsigma_{E}\rho_{RE}}{\sqrt{(p-\ell)^{2}\varsigma^{2}_{Y}+p^{2}\varsigma^{2}_{E}+2p(p-\ell)\rho_{YE}\varsigma_{Y}\varsigma_{E}}}\label{eq:rho_RYE}
\end{align}
where $\rho_{RY}$ and $\rho_{RE}$ are defined in equations \eqref{eq:rho_RY}
and \eqref{eq:rho_RE}. Using the Gaussian decomposition \eqref{eq:2N_decomposition}
with $N_{1}=N_{YE}$ and $N_{2}=N_{R}$ yields
\begin{align*}
N_{YE} & =\rho_{RYE}N_{R}+\bar{\rho}_{RYE}N^{\perp}_{R}\\
\bar{\rho}_{RYE} & :=\sqrt{1-\rho^{2}_{RYE}}=\frac{\sqrt{(p-\ell)^{2}\varsigma^{2}_{Y}\left(1-\rho^{2}_{RY}\right)+p^{2}\varsigma^{2}_{E}\left(1-\rho^{2}_{RE}\right)+2p(p-\ell)\varsigma_{Y}\varsigma_{E}\left(\rho_{YE}-\rho_{RY}\rho_{RE}\right)}}{\sqrt{(p-\ell)^{2}\varsigma^{2}_{Y}+p^{2}\varsigma^{2}_{E}+2p(p-\ell)\rho_{YE}\varsigma_{Y}\varsigma_{E}}}
\end{align*}
where $N^{\perp}_{R}$ is a standard Gaussian random variable independent
of $N_{R}$. To sum up, we have shown that
\begin{align*}
 & (p-\ell)\varsigma_{Y}N_{Y}+p\varsigma_{E}N_{E}=\\
 & \left((p-\ell)\varsigma_{Y}\rho_{RY}+p\varsigma_{E}\rho_{RE}\right)N_{R}+\\
 & \sqrt{(p-\ell)^{2}\varsigma^{2}_{Y}\left(1-\rho^{2}_{RY}\right)+p^{2}\varsigma^{2}_{E}\left(1-\rho^{2}_{RE}\right)+2p(p-\ell)\varsigma_{Y}\varsigma_{E}\left(\rho_{YE}-\rho_{RY}\rho_{RE}\right)}N^{\perp}_{R}
\end{align*}
where $N^{\perp}_{R}$ is a standard Gaussian random variable independent
of $N_{R}$. Plugging this formula into equation \eqref{eq:Clpq}
and using equation \eqref{eq:EecN} gives the formula
\end{lem}
\begin{align}
C_{\ell,p,q} & =S(t_{i})^{p}\exp\left((p-\ell)(\mu_{Y}+e^{-\kappa\Delta_{t_{i}}}r(t_{i})\Delta_{t_{i}})+p\mu_{E}\right)\nonumber \\
 & \times\exp\left(\frac{1}{2}\left((p-\ell)^{2}\varsigma^{2}_{Y}\left(1-\rho^{2}_{RY}\right)+p^{2}\varsigma^{2}_{E}\left(1-\rho^{2}_{RE}\right)+2p(p-\ell)\varsigma_{Y}\varsigma_{E}\left(\rho_{YE}-\rho_{RY}\rho_{RE}\right)\right)\right)\nonumber \\
 & \times\mathcal{E}((p-\ell)\varsigma_{Y}\rho_{RY}+p\varsigma_{E}\rho_{RE}\,,\,q)\label{eq:Clpq_explicit}
\end{align}
where $\mathcal{E}((p-\ell)\varsigma_{Y}\rho_{RY}+p\varsigma_{E}\rho_{RE}\,,\,q)$
is given explicitly in Lemma \ref{lem:Ecn}. This can be further simplified
by noticing that $\mathcal{E}(c,q)$ is always proportional to $e^{\frac{c^{2}}{2}}$,
and that $e^{-\frac{c^{2}}{2}}\mathcal{E}(c,q)$ is a polynomial in
$c$. Define
\begin{equation}
\mathcal{P}(c,q):=e^{-\frac{c^{2}}{2}}\mathcal{E}(c,q)\label{eq:Pcq}
\end{equation}
From Lemma \ref{lem:Ecn}, we know that
\begin{align*}
\mathcal{P}(c,0) & =1\\
\mathcal{P}(c,1) & =c\\
\mathcal{P}(c,2) & =c^{2}+1\\
\mathcal{P}(c,3) & =c^{3}+3c
\end{align*}
Using the notation \eqref{eq:Pcq}, $C_{\ell,p,q}$ is given by
\begin{align}
C_{\ell,p,q} & =S(t_{i})^{p}\exp\left((p-\ell)(\mu_{Y}+e^{-\kappa\Delta_{t_{i}}}r(t_{i})\Delta_{t_{i}})+p\mu_{E}\right)\nonumber \\
 & \times\exp\left(\frac{1}{2}\left((p-\ell)^{2}\varsigma^{2}_{Y}+p^{2}\varsigma^{2}_{E}+2p(p-\ell)\varsigma_{Y}\varsigma_{E}\rho_{YE}\right)\right)\nonumber \\
 & \times\mathcal{P}\left((p-\ell)\varsigma_{Y}\rho_{RY}+p\varsigma_{E}\rho_{RE}\,,\,q\right)\label{eq:Clpq_explicit_2}
\end{align}
By plugging this formula into equations \eqref{eq:BClp0}-\eqref{eq:BClp1}-\eqref{eq:BClp2}-\eqref{eq:BClp3},
we obtain that the conditional expectations $B_{\ell,p,q}$ (equation
\eqref{eq:B}) in the model \eqref{eq:dSt}-\eqref{eq:drt} are given
explicitly by
\begin{align}
B_{\ell,p,q} & =\E[\exp\left(-\ell\int^{t_{i+1}}_{t_{i}}r(\tau)d\tau\right)S(t_{i+1})^{p}r(t_{i+1})^{q}\left|\mathcal{F}_{t_{i}}\right.]\nonumber \\
 & =S(t_{i})^{p}\exp\left((p-\ell)(\mu_{Y}+e^{-\kappa\Delta_{t_{i}}}r(t_{i})\Delta_{t_{i}})+p\mu_{E}\right)\nonumber \\
 & \times\exp\left(\frac{1}{2}\left((p-\ell)^{2}\varsigma^{2}_{Y}+p^{2}\varsigma^{2}_{E}+2p(p-\ell)\varsigma_{Y}\varsigma_{E}\rho_{YE}\right)\right)\times D_{\ell,p,q}\label{eq:Blpq_final}
\end{align}
where $D_{\ell,p,q}$ is given explicitly by
\begin{align}
D_{\ell,p,0} & =1\label{eq:Dlp0}\\
D_{\ell,p,1} & =e^{-\kappa\Delta_{t_{i}}}r(t_{i})+\mu_{R}+(p-\ell)\varsigma_{R}\varsigma_{Y}\rho_{RY}+p\varsigma_{R}\varsigma_{E}\rho_{RE}\label{eq:Dlp1}\\
D_{\ell,p,2} & =\left(e^{-\kappa\Delta_{t_{i}}}r(t_{i})+\mu_{R}+(p-\ell)\varsigma_{R}\varsigma_{Y}\rho_{RY}+p\varsigma_{R}\varsigma_{E}\rho_{RE}\right)^{2}+\varsigma^{2}_{R}\label{eq:Dlp2}\\
D_{\ell,p,3} & =\left(e^{-\kappa\Delta_{t_{i}}}r(t_{i})+\mu_{R}+(p-\ell)\varsigma_{R}\varsigma_{Y}\rho_{RY}+p\varsigma_{R}\varsigma_{E}\rho_{RE}\right)^{3}\nonumber \\
 & +3\varsigma^{2}_{R}\left(e^{-\kappa\Delta_{t_{i}}}r(t_{i})+\mu_{R}+(p-\ell)\varsigma_{R}\varsigma_{Y}\rho_{RY}+p\varsigma_{R}\varsigma_{E}\rho_{RE}\right)\label{eq:Dlp3}
\end{align}
for the first few values of $q$.

\subsubsection{Closed-form formula for last time step}

Consider the final time step $[t_{N-1},t_{N}]$. Let $\Delta_{t_{N-1}}=t_{N}-t_{N-1}$.
We have
\[
W(t_{N})=\max\left(W(t_{N-1})-\pi_{N-1},0\right)\frac{S(t_{N})}{S(t_{N-1})}e^{-\alpha\Delta_{t_{N-1}}}
\]
\[
A(t_{N})=A(t_{N-1})-\pi_{N-1}
\]
\[
C_{N}(A(t_{N}))=\begin{cases}
A(t_{N}) & \mathrm{if\ }0\leq A(t_{N})\leq G_{N}\\
G_{N}+(1-\beta)(A(t_{N})-G_{N}) & \mathrm{if\ }A(t_{N})>G_{N}
\end{cases}
\]
The final payoff is
\[
\max(W(t_{N}),C_{N}(A(t_{N})))
\]
The conditional expectation of the discounted final payoff with respect
to $\mathcal{F}_{t_{N-1}}$ is given by
\begin{align*}
 & \E\left[\exp\left(-\ell\int^{t_{N}}_{t_{N-1}}r(\tau)d\tau\right)\max(W(t_{N}),C_{N}(A(t_{N})))\left|\mathcal{F}_{t_{i}}\right.\right]\\
= & \E\left[\exp\left(-\ell\int^{t_{N}}_{t_{N-1}}r(\tau)d\tau\right)\max\left(\max\left(W(t_{N-1})-\pi_{N-1},0\right)\frac{S(t_{N})}{S(t_{N-1})}e^{-\alpha\Delta_{t_{N-1}}},C_{N}(A(t_{N}))\right)\left|\mathcal{F}_{t_{i}}\right.\right]\\
= & \E\left[\exp\left(-\ell\left(Y+e^{-\kappa\Delta_{t_{N-1}}}r(t_{N-1})\Delta_{t_{N-1}}\right)\right)\right.\\
 & \left.\times\max\left(\max\left(W(t_{N-1})-\pi_{N-1},0\right)e^{-\alpha\Delta_{t_{N-1}}}\exp\left(e^{-\kappa\Delta_{t_{N-1}}}r(t_{N-1})\Delta_{t_{N-1}}+Y+E\right),C_{N}(A(t_{N}))\right)\left|\mathcal{F}_{t_{i}}\right.\right]
\end{align*}
In other words, it is of the form
\[
a\E\left[\exp\left(-\ell Y\right)\max\left(b\exp\left(Y+E\right),c\right)\right]
\]
with the $\mathcal{F}_{t_{i}}$- measurable parameters
\begin{align*}
a & =\exp\left(-\ell e^{-\kappa\Delta_{t_{N-1}}}r(t_{N-1})\Delta_{t_{N-1}}\right)>0\\
b & =\max\left(W(t_{N-1})-\pi_{N-1},0\right)e^{-\alpha\Delta_{t_{N-1}}}\exp\left(e^{-\kappa\Delta_{t_{N-1}}}r(t_{N-1})\Delta_{t_{N-1}}\right)\geq0\\
c & =C_{N}(A(t_{N}))\geq0
\end{align*}

\subsubsection{Case $b=0$}

If $b=0$, then the final conditional expectation is given by
\begin{align*}
 & a\E\left[\max\left(b\exp\left((1-\ell)Y+E\right),c\exp\left(-\ell Y\right)\right)\right]\\
= & ac\E\left[\exp\left(-\ell(\mu_{Y}+\varsigma_{Y}N_{Y})\right)\right]\\
= & ac\exp\left(-\ell\mu_{Y}+\ell^{2}\varsigma^{2}_{Y}/2\right)
\end{align*}

\subsubsection{Case $c=0$}

If $c=0$, then the final conditional expectation is given by
\begin{align*}
 & a\E\left[\max\left(b\exp\left((1-\ell)Y+E\right),c\exp\left(-\ell Y\right)\right)\right]\\
= & ab\E\left[\exp\left((1-\ell)Y+E\right)\right]\\
= & ab\exp\left((1-\ell)\mu_{Y}+\mu_{E}\right)\E\left[\exp\left((1-\ell)\varsigma_{Y}N_{Y}+\varsigma_{E}N_{E}\right)\right]\\
= & ab\exp\left((1-\ell)\mu_{Y}+\mu_{E}\right)\E\left[\exp\left(\sqrt{(1-\ell)^{2}\varsigma^{2}_{Y}+\varsigma^{2}_{E}+2(1-\ell)\rho_{YE}\varsigma_{Y}\varsigma_{E}}N_{YE}\right)\right]\\
= & ab\exp\left((1-\ell)\mu_{Y}+\mu_{E}+\left((1-\ell)^{2}\varsigma^{2}_{Y}+\varsigma^{2}_{E}+2(1-\ell)\rho_{YE}\varsigma_{Y}\varsigma_{E}\right)/2\right)
\end{align*}
where we used equation \eqref{eq:NY+NE} with $p=1$.

\subsubsection{Case $b>0$ and $c>0$}

In this case
\begin{align*}
 & a\E\left[\max\left(b\exp\left((1-\ell)Y+E\right),c\exp\left(-\ell Y\right)\right)\right]\\
= & a\E\left[\exp(\max\left(\log(b)+(1-\ell)Y+E\ ,\ \log(c)-\ell Y\right))\right]
\end{align*}
Define
\begin{align*}
X_{1} & :=\log(b)+(1-\ell)Y+E\\
 & =\log(b)+(1-\ell)\mu_{Y}+\mu_{E}+\sqrt{(1-\ell)^{2}\varsigma^{2}_{Y}+\varsigma^{2}_{E}+2(1-\ell)\rho_{YE}\varsigma_{Y}\varsigma_{E}}N_{YE}\\
 & =m_{1}+\sigma_{1}N_{YE}\\
m_{1} & :=\log(b)+(1-\ell)\mu_{Y}+\mu_{E}\\
\sigma_{1} & :=\sqrt{(1-\ell)^{2}\varsigma^{2}_{Y}+\varsigma^{2}_{E}+2(1-\ell)\rho_{YE}\varsigma_{Y}\varsigma_{E}}
\end{align*}
and 
\begin{align*}
X_{2} & :=\log(c)-\ell Y\\
 & =\log(c)-\ell\mu_{Y}+\ell\varsigma_{Y}(-N_{Y})\\
 & =m_{2}+\sigma_{2}(-N_{Y})\\
m_{2} & :=\log(c)-\ell\mu_{Y}\\
\sigma_{2} & :=\ell\varsigma_{Y}
\end{align*}
Using equation \eqref{eq:NY+NE} with $p=1$, the correlation between
the two standard Gaussian random variables $N_{YE}$ and $-N_{Y}$
is given by
\begin{align*}
\rho_{12}:= & \E\left[N_{YE}(-N_{Y})\right]\\
 & =\E\left[\frac{(1-\ell)\varsigma_{Y}N_{Y}+\varsigma_{E}N_{E}}{\sqrt{(1-\ell)^{2}\varsigma^{2}_{Y}+\varsigma^{2}_{E}+2(1-\ell)\rho_{YE}\varsigma_{Y}\varsigma_{E}}}(-N_{Y})\right]\\
 & =\frac{-(1-\ell)\varsigma_{Y}\E\left[N_{Y}N_{Y}\right]-\varsigma_{E}\E\left[N_{E}N_{Y}\right]}{\sqrt{(1-\ell)^{2}\varsigma^{2}_{Y}+\varsigma^{2}_{E}+2(1-\ell)\rho_{YE}\varsigma_{Y}\varsigma_{E}}}\\
 & =\frac{-(1-\ell)\varsigma_{Y}-\varsigma_{E}\rho_{YE}}{\sqrt{(1-\ell)^{2}\varsigma^{2}_{Y}+\varsigma^{2}_{E}+2(1-\ell)\rho_{YE}\varsigma_{Y}\varsigma_{E}}}
\end{align*}
To sum up, 
\begin{align*}
 & a\E\left[\max\left(b\exp\left((1-\ell)Y+E\right),c\exp\left(-\ell Y\right)\right)\right]\\
= & a\E\left[\exp\left(\max\left(X_{1},X_{2}\right)\right)\right]
\end{align*}
where $X_{1}\sim\mathcal{N}(m_{1},\sigma_{1})$, $X_{2}\sim\mathcal{N}(m_{2},\sigma_{2})$
and $\mathbb{C}\mathrm{orr}(X_{1,}X_{2})=\rho_{12}$. To conclude,
we use the closed-form formula for the moment generating function
of the maximum between two correlated Gaussian random variables, which
can be found for example in \citet{nadarajah2008exact}:
\begin{align}
 & \E\left[\exp\left(-\ell\int^{t_{N}}_{t_{N-1}}r(\tau)d\tau\right)\max(W(t_{N}),C_{N}(A(t_{N})))\left|\mathcal{F}_{t_{i}}\right.\right]\nonumber \\
= & a\E\left[\exp\left(-\ell Y\right)\max\left(b\exp\left(Y+E\right),c\right)\right]\nonumber \\
= & a\E\left[\exp\left(\max\left(X_{1},X_{2}\right)\right)\right]\nonumber \\
= & a\exp\!\left(m_{1}\!+\!\frac{\sigma^{2}_{1}}{2}\right)\!\times\!\Phi\!\left(\frac{m_{1}-m_{2}+\sigma^{2}_{1}-\rho_{12}\sigma_{1}\sigma_{2}}{\sqrt{\sigma^{2}_{1}+\sigma^{2}_{2}-2\rho_{12}\sigma_{1}\sigma_{2}}}\right)\!+a\exp\!\left(m_{2}\!+\!\frac{\sigma^{2}_{2}}{2}\right)\!\times\!\Phi\!\left(\frac{m_{2}-m_{1}+\sigma^{2}_{2}-\rho_{12}\sigma_{1}\sigma_{2}}{\sqrt{\sigma^{2}_{1}+\sigma^{2}_{2}-2\rho_{12}\sigma_{1}\sigma_{2}}}\right)\label{eq:final_payoff_conditional_expectation}
\end{align}
where $\Phi$ is the cumulative distribution function of the standard
Gaussian distribution.

\end{document}